\documentclass[nonblindrev]{informs3_modified}

\OneAndAHalfSpacedXI

\usepackage{natbib}
 \bibpunct[, ]{(}{)}{,}{a}{}{,}%
 \def\bibfont{\small}%
 \def\bibsep{\smallskipamount}%
 \def\bibhang{24pt}%
\TheoremsNumberedBySection  

\EquationsNumberedBySection 

\usepackage{graphics}
\usepackage{endnotes}
\usepackage{xcolor}
\usepackage{filecontents}
\usepackage{enumerate}
\usepackage{commath}
\usepackage{algorithm}
\usepackage{float}
\usepackage{algorithmic}
\usepackage{listings}
\usepackage{cancel} 
\usepackage[normalem]{ulem}
\let\footnote=\endnote

\definecolor{blue}{rgb}{0,0,0.9}
\definecolor{red}{rgb}{0.9,0,0}
\definecolor{green}{rgb}{0,0.9,0}

\def\cS{{\cal S}}

\newcommand{\R}{\mathbb R}

\newcommand{\M}{\mathbf M}
\newcommand{\V}{\mathbf V}

\newcommand{\1}{\mathbf 1}

\newcommand{\EE}{\mathbb{E}}
\newcommand{\BB}{\mathbb{B}}
\newcommand{\PP}{\mathbb{P}}

\newcommand{\cD}{\mathcal{D}}

\newcommand{\cV}{\mathcal{V}}
\newcommand{\cE}{\mathcal{E}}

\newcommand{\RR}{\mathbb{R}}

\newcommand{\cF}{\mathcal{F}}
\newcommand{\cB}{\mathcal{B}}
\newcommand{\cO}{\mathcal{O}}

\newcommand{\cC}{\mathcal{C}}
\newcommand{\cN}{\mathcal{N}}
\newcommand{\cG}{\mathcal{G}}

\newcommand{\ind}{\mathbf{1}}
\newcommand{\QED}{\ \hfill\rule[-2pt]{6pt}{12pt} \medskip}
\newcommand{\iN}{\cN^{in}}

\usepackage{amsmath}
\usepackage{algorithm}
\usepackage{algorithmic}
\usepackage{appendix}

\def \P {\mathbb{P}}
\def \R {\mathbb{R}}

\def \ind {\mathbb{I}}

\usepackage{natbib}

\bibpunct[, ]{(}{)}{,}{a}{}{,}%
\def\bibfont{\small}%
\def\bibsep{\smallskipamount}%
\def\bibhang{24pt}%
\usepackage{tcolorbox}
\usepackage{enumitem}

\usepackage{graphicx}
\usepackage{subcaption}

\usepackage[T1]{fontenc}    
\usepackage{hyperref}       
\usepackage{url}            
\usepackage{booktabs}       
\usepackage{amsfonts}       
\usepackage{nicefrac}       
\usepackage{microtype}      
\usepackage{xcolor}

\begin{document}
\RUNAUTHOR{}
\RUNTITLE{}
\TITLE{Online Learning of Independent Cascade Models with Node-level Feedback}

\ABSTRACT{
We propose a detailed analysis of the online-learning problem for Independent Cascade (IC) models under node-level feedback.   These models have widespread applications in modern social networks. Existing works for IC models have only shed light on edge-level feedback models, where the agent knows the explicit outcome of every observed edge. Little is known about node-level feedback models, where only combined outcomes for sets of edges are observed; in other words, the realization of each edge is \emph{censored}. This censored information, together with the \emph{nonlinear} form of the aggregated influence probability, make both parameter estimation and algorithm design challenging. We  establish the first confidence-region result under this setting. We also develop an online algorithm achieving a cumulative regret of $\tilde \cO(\sqrt{T})$, matching the 
theoretical regret bound for IC models with edge-level feedback. 
}

\ARTICLEAUTHORS{%
		\AUTHOR{Shuoguang Yang\thanks{Department of Industrial Engineering and Decision Analytics, The Hong Kong University of Science and Technology, Clear Water Bay, Hong Kong, Email: yangsg@ust.hk}, \ \ \ \ Van-Anh Truong\thanks{Department of Industrial Engineering and Operations Research, Columbia University, New York, NY 10027, 
 Email: vt2196@columbia.edu}  } 
}	
\maketitle

\section{Introduction}
In this paper, we focus on the node-level feedback for Independent Cascade (IC) Models \citep{kempe2003maximizing} and present a detailed analysis.  These models have been adopted widely for investigating complex networks such as social networks. 

In an IC model on a network graph, all nodes are either \emph{active} or \emph{inactive}. A node remains active once it is activated. Influence spreads over the network through a sequence of infinitely many discrete time steps.  A set of nodes called a \emph{seed set} may be activated at time step $0$. At time step $t+1$, nodes that are activated at time $t$ attempt to activate their inactive neighbors independently with particular probabilities. A node that is activated at $t$ can only make at most one attempt to activate a neighbor through an edge. This attempt occurs at time step $t+1$. If the activation succeeds, the neighbor becomes active and will attempt to activate its own neighbors at time step $t+2$. If the activation fails, this neighbor will never be activated again by this node. The IC model captures the spread of influence from a seed set over the complex network.  The spread of influence terminates when all  nodes have been activated, or all nodes have used their attempt to activate their followers.   

We consider $\cG = (\cV, \cE)$, a directed (undirected) graph.  Let $\cV$ be the set of all nodes and $\cE$ be the collection of all directed (undirected) edges. 
The IC model can be seen as equivalent to a \emph{coin flip} process \citep{kempe2003maximizing}. To be specific, the seed set $S$ is activated at the beginning of diffusion. Then,  a Bernoulli random variable $\textbf{y}(e) \sim \text{Bern}(p_{e})$ is independently sampled on each edge $e\in \cE$ with probability $p_{e}$. If we consider the subgraph consisting of edges with positive flips ($\textbf{y}(e) = 1 $) along, a node $v \in \cV \setminus S$ is activated during this diffusion process if it is connected with any node $u \in S$ through a directed path within this subgraph. 

When the influence probabilities are known, the offline solution to IC problems has been well studied \citep{kempe2003maximizing}. However, in most real-world scenarios, the influence probabilities are not readily available and must be learnt. 
Learning the diffusion probabilities is challenging. Online Influence Maximization (OIM) considers the scenario where an agent is actively selecting seed sets, observing diffusion outcomes, and learning the influence probabilities simultaneously in multiple learning rounds. In the literature, most works assume the access to the realizations of edge activations during the diffusion process \citep{wen2017online,OIM}, called \emph{edge-level feedback}. In this case, for any activated node, the agent has explicit information about which neighbor triggerred this activation. Edge-level feedback is widely used in investigating social networks but turns out to be unsuitable for some important applications where edge-level information is unavailable. 
We focus on \emph{node-level feedback} models  \citep{goyal2010learning,saito2008prediction,netrapalli2012learning,vaswani2015influence}, in contrast to edge-level feedback,  where the agent can only observe the status of nodes rather than edges. 

Node-level feedback models are common in real-world applications because of the lack of edge-level information. 
For instance, consider the spread of a rumor on a social network. People might search for information on the network, then pass this information on offline or over private channels, rather than visibly over the network. To study the spread of rumors over the social network, we might only be able to find when and what a person searched, instead of precisely who influenced each person to conduct such a search.

Another example arises in  online streaming platforms, such as Netflix. A high quality movie prompts a user to  watch other films with similar characteristics. Nevertheless, due to time constraints, the user might not watch another movie immediately, even if the user is interested.  
 When the user watches it a few days later, all the related movies previously watched by the user might be attributed an influence in triggering the current selection.  Ultimately, only node-level information is available. 

Despite the importance of node-level feedback models when investigating the spread of influence, it is challenging to develop an analytical framework even to estimate influence probabilities 
in these models. To illustrate the difficulty, suppose a set of parent nodes $\cB$ attempt to activate the same follower node $v$.
Under the IC model, the aggregated activation probability can be expressed as
$$
p(\cB,v) = 1- \prod_{u\in \cB} ( 1- p_{u,v}).
$$
Unfortunately, the resulting aggregated probability is nonconvex for most of the existing family of parametric functions, including both the linear and exponential families \citep{mccullagh2019generalized}. By using the log-transformation $z_{u,v} = -\log(1-p_{u,v})$, 
\cite{vaswani2015influence} established a framework  under which the likelihood function becomes convex and the maximum likelihood estimator could be solved numerically by conducting first-order gradient descent. 
Nevertheless, this transformation does not belong to any of the above families, so that no theoretical support can be established naturally for the performance of the estimators using the existing methods established in \cite{abbasi2011improved,li2017provably}. 

  The few available existing studies of node-level feedback have focused on developing heuristics, such as Partial Credits \citep{goyal2010learning} or Expectation Maximization (EM) \citep{saito2008prediction}, to obtain estimators for the true activation probabilities. Apart from them, \cite{vaswani2015influence} proposed a gradient-descent algorithm to find the maximum likelihood estimator. However, the theoretical performance of the maximum likelihood estimator is still lacking investigation.  In particular, a confidence-region guarantee of the maximum likelihood estimator has not been established. 
  
Moreover, under node-level feedback, when more than one edge attempt to activate the same follower node,  
  we could only observe their combined effect rather than the outcome on every edge.  For this reason, in some extreme cases where multiple edges sharing a common follower tend to be activated together, classical learning algorithms such as UCB and Thompson Sampling can only find an accurate estimation of the aggregated probabilities,  rather than that of every single edge.  Meanwhile, to obtain an approximate solution, existing algorithms require the influence probabilities on every edge as inputs rather than the aggregated probabilities. Thus, well-designed learning algorithms that 
  use node-level feedback instead of edge-level information are required for such models.

In this paper, for an edge $e\in \cE$ with feature $x_e$, we assume that the influence probability takes the form $p_e = 1- \exp(-x_e^\top \theta^*)$.  This form was first  proposed in \cite{netrapalli2012learning} and is widely adopted in node-level learning \citep{vaswani2015influence}. In this scenario, the likelihood function remains concave so that a global maximizer can be obtained.  

\textbf{Contributions and Paper Organization.} 
Our contributions to node-level online learning are two-fold. 
First, given node-level observations, we are the first to show how to construct a confidence region around the maximum likelihood estimator.  This problem has not previously been tackled, because of the nonlinearity and other complex intrinsic mathematical structures of the resulting  likelihood function. 

Then, using the confidence region result, we propose an online learning algorithm that  learns influence probabilities efficiently and selects seed sets simultaneously. We prove that our algorithm achieves an $\tilde \cO(\sqrt{T})$ scaled cumulative regret, which matches the one established in edge-level feedback models \citep{wen2017online} despite having less information available.  This is also the first problem-independent regret bound for node-level IC problems.

The rest of this paper is organized into four sections. In Section \ref{sec:review}, we review the literature related to IM. In Section \ref{sec:confidence_region}, we construct a confidence region around the maximum likelihood estimator with node-level observations. In Section \ref{sec:node_learning}, we propose an online learning algorithm and establish an upper bound for the worst-case theoretical regret. We conclude this paper in Section \ref{sec:conclusion}. 

\section{Literature Reviews}\label{sec:review}
Independent Cascade (IC) and Linear Threshold (LT) \citep{kempe2003maximizing} are two models for Influence Maximization (IM) that are widely studied. Classical IM problems consider the spread of influence over a network, where a unit reward is obtained when a node is influenced. Under a cardinality constraint, IM aims at finding a seed set $S$ so that the expected reward or the expected number of influenced nodes is maximized. Although it is NP-hard to the optimal seed $S$, under both IC and LT with cardinality constraints, \cite{kempe2003maximizing} showed that the objective function is monotone and sub-modular with respect to $S$. Thus, $(1-1/e)$-approximation solutions could be efficiently found by greedy algorithms for both IC and LT models \citep{Nemhauser1978}. In addition, variants of IM models have been considered by \cite{Bharathi2007competitive,He2016RobustIM,He2018RobustIM,Chen2016RobustIM} to handle more complex real-world scenarios. 

In scenarios where the influence probabilities are unknown, OIM bandits \citep{valko2016bandits} frameworks have been proposed, where an agent needs to interact with the network actively and update the agent's estimator based on observations. Based on the observed information, OIM bandits can be classified as edge-level feedback \citep{wen2017online,OIM} where the states of both influenced edges and nodes are observed; or node-level feedback \citep{goyal2010learning,saito2008prediction,netrapalli2012learning,vaswani2015influence}  where only the states of nodes can be observed. Node-level feedback IC provides access to much less information compared with edge-level models.  Up to now, it has been challenging to estimate influence probabilities based on node-level observations accurately. Heuristic approaches such as Partial Credit \citep{goyal2010learning} and Expectation Maximization \citep{saito2008prediction} have been proposed to obtain an estimator of influence probabilities. Maximum Likelihood Estimation (MLE) has also been applied to obtain an estimator \citep{vaswani2015influence}, but no confidence region have been established for these estimators. Several fundamental problems under node-level feedback, such as the derivation of confidence region for the maximum likelihood estimator, and the design of efficient learning algorithms, have remained largely unexplored. 

Compared to an IC model, in an LT model, the activation of each component $u$ contributes a non-negative weight $w_{u,v}$ to its neighbor $v$. For a specific node $v$, once the sum of all contributions $\sum w_{u,v}$ exceeds a uniformly drawn threshold $\delta_v \in [0,1]$, the node becomes active. Thus, under an LT model, only node-level information can be observed. 
 \cite{vaswanilearning} developed a gradient-descent approach to obtain the maximum likelihood estimator. 
 Based on the linear sum property, \cite{yang2019online} estimated the  influence probabilities by Ordinary Least Square (OLS) and analyzed its theoretical performance. They further developed an exploration-exploitation algorithm to learn the probabilities and select seed sets. \cite{li2020online} established a more potent approximation oracle to the offline LT solution and designed an UCB-type online learning algorithm. 
However, when multiple nodes $u \in \cB$ attempt to influence a specific node $v$ simultaneously, unlike the simple linear sum $ \sum_{u \in \cB} w_{u,v}$ in LT, the aggregated influence probability in IC models takes a complicated non-linear form $p(\cB,v) = 1- \prod_{u \in \cB} (1- p_{u,v})$ so that OLS is not applicable. MLE approaches have been proposed to deal with the non-linearity \citep{vaswani2015influence} but no confidence-region guarantees have yet been established. We are the first to provide a confidence-region guarantee on the maximum likelihood estimator, design online learning algorithm, and provide $\tilde \cO(\sqrt{T})$ regret under node-level feedback IC models.

Furthermore, OIM falls in the Combinatorial Multi-Armed Bandit (CMAB) regime, where each arm is a combination of multiple selected items.
 CMAB has found widespread applications in the real-world, such as assortment optimization \citep{agrawal2019mnl,agrawal2017thompson,oh2021multinomial}, and product ranking \citep{chen2020revenue}. Most CMAB models assume \emph{semi-bandit} feedback, where the agent can observe the outcome of every item within the selected arm \citep{gai2012combinatorial,chen2013combinatorial,kveton2015tight,chen2016combinatorial,Wang2017ImprovingRB}, or \emph{partial-feedback}, where the agent can only observe a subset of the selected items \citep{kveton2015combinatorial,combes2015combinatorial,katariya2016dcm,zong2016cascading,cheung2019thompson}. 
Under the  online adversarial setting, more generic models have been studied by 
\cite{mannor2011bandits, audibert2014regret}. 
In addition, OIM is closely related to Cascading Bandits, which tracks back to \cite{kveton2015cascading}, and could be treated as a special case of IC model with a chain-type graph topology.


\section{Confidence Ellipsoid under Node-level Feedback}\label{sec:confidence_region}
Due to the non-convex structure of the aggregated probability, the influence probabilities are hard to estimate under node-level feedback. 
This section starts from the MLE and converts the non-convex log-likelihood into a convex one by log-transformation. After such a transformation, 
the resulting likelihood function does not belong to any of the existing families of distributions that have been studied,  and several inherent technical challenges arise when analyzing the maximum likelihood estimator's performance. We develop a novel approach to tackle each of these challenges and build up a framework to investigate maximum likelihood estimator performance under node-level feedback.

\subsection{Maximum Likelihood Estimation}\label{sec:MLE}
Here, we analyze the MLE of the influence probabilities following the framework set up by \cite{netrapalli2012learning,vaswani2015influence}. 
For any node $v$, let $\iN(v) = \{u: (u, v) \in \cE\}$ be the collection of its parent nodes. Consider a diffusion process, node $v$ is \emph{observed} at time step $t$ if any of its parent nodes are activated at time step $t-1$, and \emph{unobserved} otherwise. 
Under a node-level feedback IC model,
at a given time step $t$, let $\cB \subset \iN(v)$ be the collection of parent nodes of $v$ activated at time step $t-1$, every node $u \in \cB$
attempts to activate $v$ simultaneously and independently, with only the status of $v$ being observable. In particular, we can  only see whether $v$ remains inactive or not. Note that when $\cB = \emptyset$, there can be no attempt to activate $v$. 

In contrast to an edge-level-feedback model, where we know the activation status of every single edge, much information is lost in the node-level-feedback model. Specifically,
if $v$ remains inactive, none of the nodes $u \in \cB$ successfully activated it.  If $v$ becomes active, we do not know precisely which node led to this activation. Thus, we can only analyze the diffusion process by seeing the combined effect of the node set $\cB$. 
We denote by $p_{\min}: = \min_{e \in \cE} p(e) >0$ the minimum activation probability over all edges in $\cG$, and denote by $\| \cdot \|$ the Euclidean-norm $\| \cdot \|_2$ unless otherwise specified.

We consider the log-transformation and define by $m(z) := 1-\exp(-z)$ for ease of notation. For every edge $e = (u,v) \in \cE$ with activation probability $p_{u,v} = 1-\exp(-z_{u,v})$, we assume that there exists an unknown vector $\theta^* \in \R^d$ such that $ z_{u,v}= x_{u,v}^\top \theta^*$, where $x_{u,v}$ is the feature this edge. 
Equivalently, we have $ p_e = 1- \exp(- z_e) = 1- \exp(-x_e^\top \theta^*)$. 
 This form of influence probability generalizes the framework set up in \cite{netrapalli2012learning,vaswani2015influence}. Meanwhile, when the influence probabilities are small, it is easy to see that $p_e = 1-\exp(-x_e^\top\theta^*)\approx x_e^\top \theta^*$ so that it is a good approximation of the linear influence probability $p_e = x_e^\top\theta^*$ considered in \cite{wen2017online}. 



We index every edge $e \in \cE$ as $e_1, \cdots, e_{|\cE|}$. Note that in the \emph{tabular} case where edge features are unknown, we denote by $d_k$ the in-degree of the follower node that edge $e_k$ is pointing to,
and then set $x_{e_k} = (\underbrace{0,0,\cdots,0}_{k-1\,\,0' s},1/d_k,0,\cdots, 0) \in \R^{|\cE|}$. 
That is, the $k$-th components of $x_{e_k}$ is $1/d_k$ and other components are zero.
For any node set $\cB$, let $x(\cB,v) := \sum_{u \in \cB} x_{u,v}$, by setting edge features in this default way,  it ensures that 
$\| x(\cB,v) \|\leq \sum_{u \in \cB} \| x_{u,v}\| \leq 1$. Then, we can derive a tractable maximum likelihood estimator for the influence probabilities of every edge in the absence of any given feature information on the edges.

Given any node set $\cB$, we have 
\begin{align*}
p(\cB,v) :=  1-\prod_{u \in \cB} \Big ( 1- m(x_{u,v}^\top \theta^*) \Big ) = 1- \exp \Big (-\sum_{u \in \cB} x_{u,v}^\top \theta^* \Big)  = m \Big (x(\cB,v)^\top \theta^* \Big).
\end{align*}
Thus, the effect of node set $\cB$ can be equivalently expressed as that of a \emph{hyper-edge} with feature $x(\cB,v)$. For ease of presentation, we adopt this hyper-edge notation throughout this paper.  

Given the observations of $n$ hyper-edges $x_1,\cdots, x_n$, we denote by $Y_1,\cdots, Y_n$ their corresponding realizations, where $Y_i = 1$ if $x_i$ successfully activates its follower node, and $Y_i = 0$ otherwise. The likelihood function for the sequence of events can be written as: 
$$
L_n(\theta;Y_1,\cdots, Y_n) = \prod_{i=1}^n m(x_i^\top\theta)^{Y_i} \Big (1- m(x_i^\top\theta)\Big )^{1- Y_i},
$$
with the log-likelihood function being 
\begin{equation}\label{eq:log-likelihood}
l_n(\theta;Y_1,\cdots, Y_n) = \sum_{i=1}^n \Big ( Y_i \log m(x_i^\top\theta) +(1- Y_i) \log (1- m(x_i^\top\theta))\Big ).
\end{equation}
Taking derivatives with respect to $\theta$, we have
\begin{equation}\label{eq:gradient}
    \nabla l_n(\theta;Y_1,\cdots, Y_n) = \sum_{i=1}^n x_i (\frac{Y_i}{m(x_i^\top\theta)} - 1),
\end{equation}
and 
\begin{equation}\label{eq:hessian}
    \nabla^2 l_n(\theta;Y_1,\cdots, Y_n) = -\sum_{i=1}^n x_i x_i^\top \frac{Y_i \exp(-x_i^\top\theta)}{m^2(x_i^\top\theta)}.
\end{equation}
 Here, the Hessian is negative semi-definite, which implies the concavity of the log-likelihood function. Hence, first-order methods such as gradient ascent lead to convergence of the maximum likelihood estimators.  We adopt the gradient ascent method to derive the maximum likelihood estimator of diffusion probabilities.

\subsection{Assumptions and Preliminaries}
Learning the diffusion probabilities for an IC model is challenging, especially when the agent can only make observations at the node level. Existing works have focused on developing heuristics to obtain an estimator for the true activation probability \citep{goyal2010learning,saito2008prediction,vaswani2015influence}.
Furthermore, because of the nonlinear aggregated influence probability that IC models preserve, the OLS approach adopted in LT models cannot be applied.  A new method for learning the diffusion probabilities must be developed instead.


 Throughout this paper, we impose the following normalization assumption on all possible combinations of hyper-edges. 
\begin{assumption}\label{assumption:feature_norm}
For every $v \in \cV$ and every $\cB \subset \cN^{in}(v)$, $\| \sum_{u\in \cB} x_{u,v}\|\leq 1$. 
\end{assumption}
Note that this assumption can always be satisfied by re-scaling on the feature space, and also holds under our setting of edge features in Section \ref{sec:MLE} for the tabular case in the absence of feature information.
Let $\BB_1(\theta^*) := \{\theta: \| \theta - \theta^*\| \leq 1 \}$ denote the unit ball around $\theta^*$.
Let $\kappa > 0 $ be the minimum value of $\exp(-x^\top\theta)/m^2(x^\top\theta)$ over the unit ball $\BB_1(\theta^*)  := \{\theta: \| \theta - \theta^*\| \leq 1 \}$ around $\theta^*$ for any $x \in \RR^{d}$ such that $\|x \|\leq 1$.  That is, 
\begin{equation}\label{def:kappa}
  \kappa:= \inf \Big \{ \frac{\exp(-x^\top\theta)}{m^2(x^\top\theta)} \Big | \|\theta - \theta^* \|\leq 1 , x\in \RR^d,\| x\|\leq 1 \Big \}.  
\end{equation}
Note that $\kappa >0$ holds by the fact that $\exp(-x^\top\theta)/m^2(x^\top\theta) >0$ for all $\| \theta \| < \infty$ and $\|x \|\leq 1$. Then for all hyper-edge $e$ with feature $x_e$, under Assumption \ref{assumption:feature_norm} that $\|x_e \| \leq 1$, it is easy to see $\exp(-x_e^\top\theta)/m^2(x_e^\top\theta) \geq \kappa$.

Recall the log-likelihood \eqref{eq:log-likelihood}, its first-order derivative \eqref{eq:gradient}, and Hessian \eqref{eq:hessian}. 
Clearly, in the first-order derivative \eqref{eq:gradient}, a node provides information about the influence probabilities as long as it is observed. Nevertheless, one can also see that only data from hyper-edges that successfully activate their follower nodes are included in the Hessian \eqref{eq:gradient}.
Let $\hat \theta_n$ be the maximum likelihood estimator to \eqref{eq:log-likelihood},
to better investigate its performance, we
denote by
\begin{equation}\label{def:V}
 \V_n := \sum_{i=1}^n x_i x_i^\top Y_i,\   \text{ and } \ \M_n := \sum_{i=1}^n x_i x_i^\top. 
\end{equation}
Here, $\V_n$ and $\M_n$ are two positive semi-definite matrices containing successful activations only, and all activations, respectively, so that both are essential in establishing confidence regions as we will discuss presently.

\textbf{Importance of $\V_n$.}
First, for $\theta \in \BB_1(\theta^*) $, we have 
$ - \nabla^2 l(\theta)  \succeq \sum_{i=1}^n x_i x_i^\top Y_i \kappa =  \kappa \V_n$. Thus, it is possible to provide a guarantee of the local concavity of the log-likelihood function \eqref{eq:log-likelihood} around the estimator $\hat \theta_n$ in terms of $\V_n$.  The convexity, in turns, determines the ``size'' of the confidence region around the estimator.

To do this, for any vector $z \in \R^d$ and symmetric matrix $V \in \R^{d \times d}$, we define a $V$-norm  as $\|z \|_{V} = \sqrt{z^\top V z}$.  Based on the form of the Hessian derived \eqref{eq:hessian}, it is natural to assess the difference $\hat \theta_n - \theta^*$ by whether or not $\theta^*$ falls in the region $\cC_n = \{\theta: \|\theta - \hat \theta_n\|_{\V_n}  \leq c_n \}$ with a certain probability.  By the eigenvalue decomposition, it is easy to see that $\cC_n$ is contained in an Euclidean ball with radius $r = c_n/\sqrt{\lambda_{\min}(\V_n)}$, namely $\BB_r(\hat \theta_n) = \{ \theta : \|\theta - \hat \theta_n\|^2 \leq c_n^2/\lambda_{\min}(\V_n)\}$.  Instead of using this standard ball, we find it convenient to use the $\V_n$-norm ball.  The latter is an ellipsoid and provides better estimation in the directions with larger eigenvalues.

\textbf{Importance of $\M_n$.}
Considering the $\V_n$ distance alone is insufficient to measure $\hat \theta_n - \theta^*$ accurately. By definition \eqref{def:V}, $\V_n = \sum_{i=1}^n x_i x_i^\top Y_i$ is the collection of feature information from successful edges, which involves randomness due to the cascade process. To construct a $\V_n$-ball as the confidence ellipsoid, we must investigate the relationship between $\V_n$ and its expectation. Consider the expectation $\EE[\V_n]$, we have $\EE[\V_n] = \EE [ \sum_{i=1}^n x_i x_i^\top Y_i] = \sum_{i=1}^n x_i x_i^\top  p_i \succeq p_{\min} \sum_{i=1}^n x_i x_i^\top = p_{\min} \M_n$,
 with $p_{\min}: = \min_{e \in \cE} p(e)$ being the minimal activation probability over all edges. We conclude that $\EE[\V_n] \succeq p_{\min} \M_n$. In this way, $\M_n$ is linked with the expectation of $\V_n$.
 

\textbf{Bridge connecting $\M_n$ and $\V_n$.}
As stated,
investigating the relationship between $\V_n$ and $\M_n$ is crucial to our confidence region analysis. In most of the existing literature, such as \cite{li2017provably}, it has not been necessary to explore this relationship.  The reason is that in past models, $\V_n$ is defined as $\V_n := \sum_{i=1}^n x_i x_i^\top$, unlike the IC model where $\V_n = \sum_{i=1}^n x_i x_i^\top Y_i$. Given the observed edges $x_1,\cdots, x_n$, in their models, $\V_n$ is a deterministic quantity so that $\V_n = \EE[\V_n]$.  Thus it is unnecessary to explore $\V_n$ and its expectation, making their analysis much less challenging.

To relate $\M_n$ and $\V_n$, we consider the following Semi-Definite Program (SDP): 
\begin{equation} \label{prob:sdp_rho}
        \rho_n^* : = \max \{ \rho | \V_n - \rho \M_n \succeq 0 \},
\end{equation}
in order to obtain $\M_n \succeq \V_n \succeq \rho_n^* \M_n $. With such a relationship, $\lambda_{\min}(\M_n)$ and $\rho_n^* \lambda_{\min}(\M_n)$ serve as the upper and lower bounds for $\lambda_{\min}(\V_n)$, respectively, which play substantial roles in the forthcoming analysis. 

\textbf{Sub-Gaussian randomness in edge realizations.}
Finally, for ease of notation, we denote by $\epsilon_i = Y_i - m(x_i^\top\theta^*)$. That is, 
\begin{equation*}
\epsilon_i = \begin{cases}
1- m(x_i^\top\theta^*) & \text{ with prob. } m(x_i^\top\theta^*), \\
-m(x_i^\top\theta^*) & \text{ with prob. } 1- m(x_i^\top\theta^*).
\end{cases}
\end{equation*}
It is easy to see that $\epsilon_i$ is a sub-Gaussian random variable such that $ \EE[ \epsilon_i] = 0$. In particular, for any $s > 0$, we have 
\begin{equation}\label{eq:subGaussian}
    \EE[\exp(s \epsilon_i)] \leq \exp(\frac{s^2}{2}).
\end{equation}

\subsection{Confidence Ellipsoid}
In the rest of this section, we show how to construct a confidence ball around the maximum likelihood estimator $\hat \theta_n$. Building up confidence balls for maximum likelihood estimators has drawn many researchers' attention because of its widespread application in various practical problems. 
When the observations are independent and identically distributed, Fisher information has been considered to be one of the most efficient tools to provide a lower bound on an unbiased estimator's variance, which further leads to a confidence region with a rather succinct form \citep{Fisher001OA}. When the observations are dependent, which is the case for most stochastic bandit problems, it becomes hard to analyze the estimator's performance.
 \cite{abbasi2011improved} considered the linear stochastic bandit problem and addressed the dependent observation issue
using a martingale approach.  They constructed a sharp confidence ball in all directions of the feature space. \cite{li2017provably} adopted generalized linear models (GLM) and extended the linear results to an  exponential family of functions in logistic and probit regression. However, the likelihood function for node-level IC assumes a more complex structure than that found in all existing works, so that we can apply none of these approaches directly to this case. 
The performance of the estimator $\hat \theta_n$ is therefore worthy of further study here.

\textbf{Comparisons with closest works.}
Let $\mu(x_i^\top\theta)$ be any non-negative function representing the activation probability for a random variable with feature $x_i$ (similar to $m(x_i^\top\theta)$ here).  \cite{li2017provably}, the closest to our work,  considered a log-likelihood function whose derivative takes the form 
\begin{equation*}
    \nabla l_n(\theta) = \sum_{i=1}^n x_i \Big (\mu(x_i^\top\theta) - Y_i \Big ) = \sum_{i=1}^n x_i \Big (\mu(x_i^\top\theta) -\mu(x_i^\top\theta^*) -\epsilon_i \Big ).
\end{equation*}
Our work differs from that of \cite{li2017provably} in three respects, which we will outline here.  
\begin{itemize}
    \item 
First, the Hessian in the case considered by \cite{li2017provably} is  $\nabla^2 l_n(\theta)  = \sum_{i=1}^n x_i x_i^\top \dot{\mu}(x_i^\top\theta)$, which implies $\V_n = \M_n$. Thus, as we mentioned earlier, they do not need to investigate the relationship between $\V_n$ and $\M_n$, which significantly simplifies their analysis.  

\item Secondly, let $G_n(\theta) :=  \sum_{i=1}^n x_i \Big (\mu(x_i^\top\theta) - \mu(x_i^\top\theta^*) \Big )$. 
Clearly, $G_n(\theta^*) = 0$ and $G_n( \hat  \theta_n) = \sum_{i=1}^n \epsilon_ix_i$. Using the fact that $G_n(\hat  \theta_n)$ is a sum of independent bounded sub-Gaussian random vectors, \cite{li2017provably} established  that $\theta^*$ falls in the ellipsoid $(\theta -  \hat  \theta_n)^\top\M_n (\theta - \hat  \theta_n) \leq c^2$ with a certain probability, where $\M_n = \sum_{i=1}^n x_i x_i^\top$ and $c>0$ is a constant.
However, the above method does not directly apply to our case. If we follow that approach, after decomposing $Y_i = m(x_i^\top\theta^*) + \epsilon$, we have $\nabla l_n(\theta) = \sum_{i=1}^n x_i (\frac{m(x_i^\top\theta^*) + \epsilon_i}{m(x_i^\top\theta)} - 1)$. 
Consider the corresponding link function $ G_n(\theta) = \sum_{i=1}^n x_i (\frac{m(x_i^\top  \theta^*)}{m(x_i^\top\theta)} - 1)$.  We obtain $G_n(\theta^*) = 0$ and $G_n( \hat  \theta_n) = \sum_{i=1}^n x_i \epsilon_i/m(x_i^\top\hat   \theta_n)$. It is important to note that $\hat  \theta_n$ is dependent on the realization of $\epsilon_i$'s. Thus, $\epsilon_i / m(x_i^\top \hat  \theta_n)$ is no longer a sub-Gaussian random variable, and $\EE[  G_n(\hat \theta_n)]= 0$ may not hold anymore.  Thus, the approach of \cite{li2017provably} cannot be directly applied to our problem. 

\item Finally, consider the confidence ball.  In contrast to the case in \cite{li2017provably},  $\V_n = \sum_{i=1}^n x_i x_i^\top  Y_i $ is correlated with the realizations of $Y_i$'s and thus the value of derived $\hat \theta_n$ in our problem, as both $\V_n$ and $ \hat  \theta_n$  depend on the history of observations.  The randomness of $\V_n $ leads to a more  complex analysis as well.
\end{itemize}
\textbf{Confidence $\V_n$-ellipsoid.}
To solve our problem, we propose a novel method that avoids using a link function $G(\hat \theta_n)$ with non-sub-Gaussian terms $\epsilon_i/m(x_i^\top \hat \theta_n)$.  Instead, we investigate more deeply the relationship between $\V_n$ and $\hat \theta_n$. Our result is as follows.
\begin{theorem}\label{thm:confidence_ball}
Suppose Assumption \ref{assumption:feature_norm} holds. Suppose there exist $\kappa >0$ satisfying \eqref{def:kappa} and  $\rho >0$ such that $\V_n \succeq \rho \M_n \succ 0$, and let $R := \max_{e \in \cE} 1/p(e)$. For any $\delta_1 >0$ such that $\kappa^2  \rho^2 \lambda_{\min}(\M_n) \geq 16R^2\Big ( d +\log(1/\delta_1) \Big )$, 
we have 
 \begin{equation} \label{eq:confidence_ball_1}
  \| \hat \theta_n - \theta^*\|_{\V_n}^2 \leq \frac{4R^2}{\kappa^2 \rho} \Big ( \frac{d}{2} \log(1 + 2n/d) + \log (1/\delta_2)\Big ),
 \end{equation}
 with probability at least $1-\delta_1 - \delta_2$.
\end{theorem}
\proof{Proof.} To start with, we recall the  first order derivative \eqref{eq:gradient} 
\begin{equation*}
    \nabla l_n(\theta) = \sum_{i=1}^n x_i (\frac{Y_i}{m(x_i^\top \theta)} - 1).
\end{equation*}
By setting it to zero, the maximal likelihood estimator $\hat \theta_n$ satisfies the following equation
\begin{equation}\label{eq:theta_estimator}
   \nabla l_n(\hat \theta_n)  =  \sum_{i=1}^n x_i (\frac{Y_i}{m(x_i^\top \hat \theta_n)} - 1) = 0.
\end{equation}
It is also easy to see 
\begin{equation}\label{eq:first_order_theta_ast}
\nabla l_n(\theta^*) = \sum_{i=1}^n x_i (\frac{Y_i}{m(x_i^\top  \theta^*)} - 1).
\end{equation}
Consider the unit ball $\BB_1(\theta^*) = \{ \theta : \| \theta - \theta^*\| \leq 1 \}$ around $\theta^*$, our analysis relies on the event that $\{ \hat \theta_n \in \BB_1(\theta^*) \}$. In what follows, we first characterize the probability that $\hat \theta_n \in \BB_1(\theta^*)$, and then investigate the performance of $\hat \theta_n$ under such condition. 

For any $\theta \in \BB_1(\theta^*)$, by the Mean Value Theorem,  there exists $\bar \theta := q \theta + (1-q) \theta^*$ for $0 < q < 1$ such that 
\begin{equation*}
    \nabla l_n(\theta) - \nabla l_n(\theta^*) = \sum_{i=1}^n x_iY_i (\frac{1}{m(x_i^\top \theta)}-\frac{1}{m(x_i^\top \theta^*)}) = \nabla^2 l_n(\bar \theta)(\theta - \theta^*).
\end{equation*}
Since $\bar \theta = q \theta + (1-q) \theta^*$ is a convex combination of $ \theta$ and $\theta^*$, $\bar \theta$ falls in the unit ball $ \BB_1(\theta^*)$ as well. By the definition of $\kappa$ \eqref{def:kappa}, we have $\exp(-x_i^\top \bar \theta)/ m^2(x_i^\top \bar \theta) \geq \kappa >0$  for $i = 1,\cdots,n$, and  $\V_n = \sum_{i=1}^n Y_i  x_i x_i^\top $, which further implies 
$$
-\nabla^2 l_n(\bar \theta) = \sum_{i=1}^n x_i x_i^\top  Y_i \frac{\exp(-x_i^\top \bar \theta)}{ m^2(x_i^\top \bar \theta)} \succeq \sum_{i=1}^n x_i x_i^\top  Y_i \kappa = \kappa \V_n \succ 0_{d\times d} .
$$
It is easy to see $(\theta - \theta^*)^\top (\nabla l_n(\theta) - \nabla l_n(\theta^*)) > 0$ for any $\theta \neq \theta^*$ so that $\nabla l_n(\theta)$ is an injection from $\R^d$ to $\R^d$. 
Define $H_n(\theta) := \|\nabla l_n(\theta) - \nabla l_n(\theta^*) \|_{\V_n^{-1}}^2 $, we have $H_n(\theta^*) = 0$, and   
\begin{align*}
 H_n(\theta) 
    & =  (\theta - \theta^*)^\top \nabla^2 l_n(\bar \theta) \V_n^{-1} \nabla^2 l_n(\bar \theta) (\theta - \theta^*)\\
&    \geq  \kappa^2( \theta - \theta^*)^\top \V_n (  \theta - \theta^*)\\
&     \geq \kappa^2  \lambda_{\min}(\V_n) \| \theta - \theta^*\|^2 \geq \kappa^2 \rho \lambda_{\min}(\M_n) \| \theta - \theta^*\|^2,
\end{align*}
where we use the fact that $\nabla^2 l_n(\bar \theta) \succeq \kappa \V_n \succ 0$ in the first inequality and $\V_n \succeq \rho \M_n$ in the last one.
Clearly, $H_n(\theta) \geq \kappa^2 \rho \lambda_{\min}(\M_n)$ for $\| \theta - \theta^* \| = 1$.
As $H_n(\cdot)$ is a continuous function and $H_n(\theta^*) = 0$, by basic topology, we have 
$$
\left \{\theta \Big | H_n(\theta) \leq   \kappa^2 \rho \lambda_{\min}(\M_n) \right \} \subset \BB_1(\theta^*).
$$
That is, for any $\theta $ such that $H_n( \theta) \leq \kappa^2 \rho \lambda_{\min}(\M_n)$, it is contained in the unit ball $ \BB_1(\theta^*)$. 
As we aim to characterize the behavior of $\hat \theta_n$, our first target is to show $H_n(\hat \theta_n) < \kappa^2 \rho \lambda_{\min}(\M_n)$ with high probability, which leads to $\| \hat \theta_n - \theta^*\| \leq 1$.
\\
\textbf{Bound for $\P(\| \hat \theta_n - \theta^*\| \leq 1)$: }
Consider $H_n(\hat \theta_n)$. Since $\nabla l_n(\hat \theta_n) = 0$, it is easy to see $H_n(\hat \theta_n) = \|\nabla l_n(\hat \theta_n) - \nabla l_n(\theta^*) \|_{\V_n^{-1}}^2 = \| \nabla l_n(\theta^*) \|_{\V_n^{-1}}^2$. Using the condition $\V_n \succeq \rho \M_n \succ 0$, we have 
 $$H_n( \hat \theta_n) = \|\nabla l_n(\theta^*) \|_{\V_n^{-1}}^2   = \nabla l_n(\theta^*)^\top  \V_n^{-1} \nabla l_n(\theta^*) \leq \nabla l(\theta^*)^\top  (\rho \M_n)^{-1} \nabla l_n(\theta^*)  =  \|\nabla l_n(\theta^*) \|^2_{\M_n^{-1}} /\rho. $$
 Suppose there exists an intermediate term $U_n$  such that $ \|\nabla l_n(\theta^*) \|^2_{\M_n^{-1}} \leq U_n \leq \kappa^2 \rho^2 \lambda_{\min}(\M_n)$, then we obtain  
 \begin{equation}\label{eq:H_n}
  H_n( \hat \theta_n) \leq \frac{\|\nabla l(\theta^*) \|^2_{\M_n^{-1}} }{\rho} \leq \frac{U_n }{\rho} \leq \kappa^2 \rho \lambda_{\min}(\M_n),    
 \end{equation}
which further implies $\| \hat \theta_n - \theta^* \| \leq 1$. 
We then show the existence of such an $U_n$ in the following analysis. 

Recall that 
$Y_i = m(x_i^\top \theta^*) + \epsilon_i$, 
where $\epsilon_i$ is a zero-mean bounded sub-Gaussian random variable such that $\epsilon_i = 1- m(x_i^\top \theta^*)$ with probability $m(x_i^\top \theta^*)$ and $\epsilon_i = - m(x_i^\top \theta^*)$ with probability $1- m(x_i^\top \theta^*)$. Let $v_i := \frac{Y_i}{m(x_i^\top \theta^*)} - 1$.  Then $v_i$ can be equivalently expressed as 
$$v_i = \frac{m(x_i^\top \theta^*) + \epsilon_i}{m(x_i^\top \theta^*) }- 1 = \frac{\epsilon_i}{ m(x_i^\top\theta^*)}. $$
Clearly, $v_i$ is a zero-mean bounded $1/m(x_i^\top\theta^*)$-sub-Gaussian random variable. Let $R = \max_{e \in \cE} 1/m(x_e^\top \theta^*) = 1/p_{\min}$.  By  \eqref{eq:subGaussian}, we have 
for any $s>0$,
\begin{equation*}
    \EE[\exp(s v_i)] \leq \exp(\frac{s^2 R^2}{2}), \,\,\,\text{ for }i=1,\cdots, n.
\end{equation*}
Then, by recalling \eqref{eq:first_order_theta_ast}, it can be seen that 
$\nabla l_n(\theta^*) = \sum_{i=1}^n x_i (\frac{Y_i}{m(x_i^\top \theta^*)} - 1)  = \sum_{i=1}^n x_i v_i $ is also sum of $R$-subgaussian random vectors. To ensure $\|  \hat \theta_n - \theta^*\| \leq 1$,
a bound on $\| \nabla l_n(\theta^*)  \|_{\M_n^{-1}}^2$ is provided by Lemma \ref{lemma:loose_bound} in Appendix Section \ref{app:B5}  for $R$-subgaussian random $v_i$'s. In particular, with probability at least $1-\delta_1$, we have 
\begin{equation*}
\| \nabla l_n(\theta^*)  \|_{\M_n^{-1}}^2  = \| \sum_{i=1}^n x_i v_i \|_{\M_n^{-1}}^2 \leq 16R^2 \Big ( d +\log(1/\delta_1) \Big ).
\end{equation*}

Clearly, by choosing any $\delta_1 > 0 $ such that $\kappa^2  \rho^2 \lambda_{\min}(\M_n) \geq 16R^2\Big ( d +\log(1/\delta_1) \Big )$, and setting $U_n = 16R^2\Big ( d +\log(1/\delta_1) \Big )$ in \eqref{eq:H_n}, 
we obtain  
\begin{equation}\label{eq:unit_prob_bound}
\P \Big ( \| \hat \theta_n - \theta^*\| \leq 1  \Big ) \geq 1-\delta_1.    
\end{equation}
\textbf{Confidence $\V_n$-ellipsoid:}
Under the scenario  $\| \hat \theta_n - \theta^*\| \leq 1$, we have
\begin{equation}\label{eq:confidence_region_VM}
  \|  \hat \theta_n - \theta^*\|_{\V_n}^2 \leq \frac{H_n(\hat \theta_n)}{\kappa^2}= \frac{\| \nabla l_n(\theta^*) \|_{\V_n^{-1}}^2}{\kappa^2} \leq \frac{\|\nabla l_n(\theta^*) \|^2_{\M_n^{-1}} }{ \kappa^2 \rho}.  
\end{equation}
Next, we focus on the case where $\|  \theta_n - \theta^*\|^2 \leq 1$   and provide a bound for  $\|\nabla l_n(\theta^*) \|^2_{\M_n^{-1}}$. When $\lambda_{\min}(\M_n) \geq 1$, by using Lemma \ref{lemma:tight_bound} in Appendix Section \ref{app:B5} for a general $R$, we have with probability at least $1-\delta_2$,
$$
\| \nabla l_n(\theta^*) \|_{\M_n^{-1}}^2 = \| \sum_{i=1}^n x_i v_i \|_{\M_n^{-1}}^2 \leq 4R^2\Big ( \frac{d}{2} \log(1 + 2n/d) + \log (1/\delta_2)\Big ).
$$
 Putting the above inequality into \eqref{eq:confidence_region_VM}, and applying \eqref{eq:unit_prob_bound}, then for any $\delta_1 > 0 $ such that $\kappa^2  \rho^2 \lambda_{\min}(\M_n) \geq 16R^2\Big ( d +\log(1/\delta_1) \Big )$, we obtain 
 \begin{equation*}
    \|\hat \theta_n - \theta^*\|_{\V_n}^2 \leq \frac{4R^2}{\kappa^2 \rho} \Big ( \frac{d}{2} \log(1 + 2n/d) + \log (1/\delta_2)\Big )
 \end{equation*}
 with probability at least $1-\delta_1 - \delta_2$, which completes the proof. 
 \QED

In the above theorem, we establish a confidence ball around the maximum likelihood estimator $\hat \theta_n$ given that $\V_n \succeq \rho \M_n$, where $\rho$ is derived by solving the SDP \eqref{prob:sdp_rho}. To the best of our knowledge, this is the first confidence region result for node-level IC models, which provides a theoretical guarantee on the performance of maximum likelihood estimators and preserves substantial importance for real-world applications.

\section{Online Learning with Node-level Feedback} \label{sec:node_learning}
In Section \ref{sec:confidence_region}, we  establish a confidence region for the derived maximum likelihood estimator, which could  apply directly to offline parameter estimation. Despite that result, in real-world scenarios the data are not typically available as a prior and can only be collected by making decisions and interacting repeatedly with the environment. 
In this section, we consider OIM under node-level feedback, and propose an online learning algorithm that actively learns the influence probabilities and selects seed sets simultaneously in multiple rounds. In this section, we propose an online learning algorithm and derive its theoretical worse-case regret guarantee. 

\subsection{Challenges in Algorithm Design and Analysis:} \label{sec:challenges}
Despite 
the confidence region established in Section \ref{sec:confidence_region}, it remains unclear how to design an efficient online learning algorithm and analyze its theoretical performance because of the three significant challenges below:
\begin{itemize}
    \item First, the classical UCB approaches do not apply directly  to node-level feedback settings. Briefly speaking, in each learning round, classical UCB-type algorithms construct an \emph{optimistic estimator} of the influence probability for each edge and select a seed set using the optimistically estimated influence probabilities. With edge-level feedback, each edge's estimators are updated whenever observed and become more accurate with more observations.  Nevertheless, the information structure is different under node-level feedback. In particular, as the agent can only observe the combined outcome for a set of edges, the agent's optimistic estimators for every single edge are not necessarily improved using node-level information. 
    
    \begin{figure}
    \centering
    \includegraphics[scale=0.7]{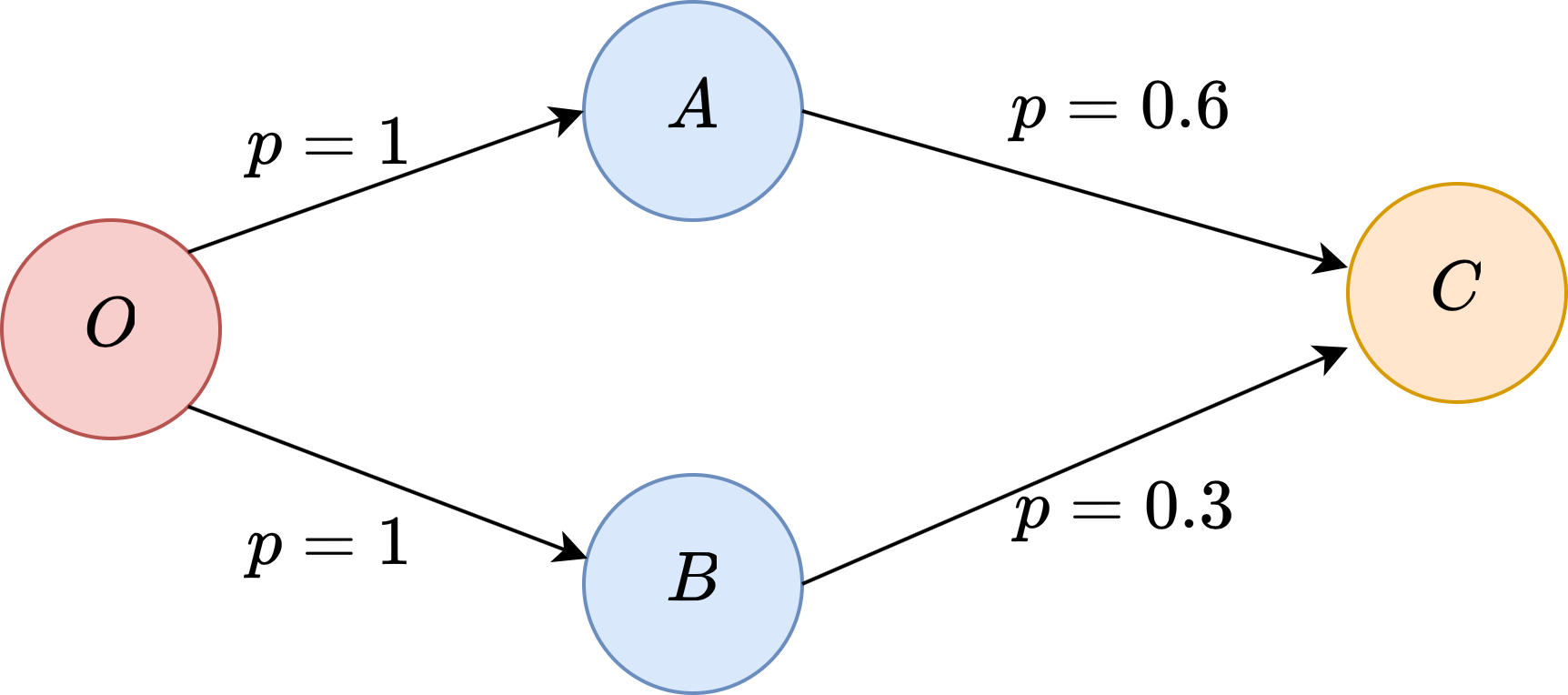}
    \caption{An example where individual optimistic estimators cannot be improved by UCB-Type approaches.
    }
    \label{fig:1}
\end{figure}

    To illustrate this, we consider a simple directed graph consisting of four nodes, $O, A, B$, and $C$, and four directed edges $(O,A), (O,B), (A,C)$, and $(A,B)$ shown in Figure \ref{fig:1}, with $p_{O,A} =  p_{O,B} = 1, p_{A,C} = 0.6$, and $p_{B,C} = 0.3$. When node $O$ is selected as the seed set, nodes $O$ and $A$ would always be activated simultaneously at time step $t=1$. Subsequently, $A, B$ would attempt to influence their inactive neighbor $C$, with probability $p_{A,C} = 0.6$ and $p_{B,C} = 0.3$, respectively. Notably, under node-level feedback, only the combined effect of both $A$ and $B$ rather than the realizations on each edges $(A,C)$ and $(B,C)$ are observed. Suppose we were to apply classical UCB-type approaches to design online algorithms.  Optimistic estimators are required for each edge. Nevertheless, using node-level information alone, the optimistic estimators could only be improved to their true \emph{aggregated} probability, $u_{A,C} = u_{B,C} = 0.72$, even with infinite observations.  These bounds are far from the true \emph{individual} influence probabilities.
Without better knowledge of the optimistic estimators, the selected seed set will not necessarily improve, making UCB approaches inapplicable.
    For this reason, we need to design alternative algorithms to overcome this issue significantly. 
    
    \item The second challenge  is the unknown distribution of $\rho_n^*$ involved in the confidence region established by Theorem \ref{thm:confidence_ball}. Recall that $\rho_n^*$ is the optimal solution to Prob.\eqref{prob:sdp_rho}. In particular, we have
$ \rho_n^* = \max \{ \rho | \V_n - \rho \M_n \succeq 0 \}$ 
and $\M_n \succeq \V_n \succeq \rho_n^* \M_n $.  
Clearly, as both $\V_n$ and $\M_n$ are positive semi-definite matrices, $\rho_n^* \geq 0$ always holds. However, the trivial solution $\rho_n^* = 0$ to \eqref{prob:sdp_rho} does not shed light on the relationship between $\V_n$ and $\M_n$. Moreover, let  $\cC_n$ be the confidence region constructed in  \eqref{eq:confidence_ball_1}, as the radius of this ellipsoid is proportional to $1/\rho_n^*$, we must investigate the distribution of $\rho_n^*$ to ensure the boundedness of the confidence region. 

\item Third, the confidence region $\cC_n$ \eqref{eq:confidence_ball_1}  is constructed using feature information from \emph{activated} hyper-edges, in particular, $\V_n$, instead of the \emph{full} information matrix $\M_n$. Existing approaches in the bandits literature can only analyze the regret when we build up the confidence region $\cC_n$ using the full information matrix $\M_n$ instead of $\V_n$. It remains unclear how to investigate the performance of the online algorithm in the latter scenario. 
\end{itemize}
These challenges mentioned above make node-level IC learning highly nontrivial. To address these issues, we develop a novel learning algorithm that consists of exploration and exploitation phases and study its theoretical performance.

\subsection{Assumptions and Preliminaries}
Recall the Maximum Likelihood Estimation in Section \ref{sec:confidence_region},  for each edge $e = (u,v) \in \cE$ with activation probability $ p_{u,v}$, by taking the log-transformation $ x_{u,v}^\top \theta^* = -\log (1-  p_{u,v})$, the activation probability on each edge can be expressed as $p_{u,v} = 1- \exp(- x_{u,v}^\top \theta^*)$. 
Also, we define $p(\theta)$ as the collection of influence probabilities under parameter $\theta$ such that $p_e(\theta) = 1-\exp(x_e^\top  \theta)$ for all $e\in \cE$. In particular, we have the true influence probability $ p_e = p_e(\theta^*)$. 

We first introduce some mild assumptions on the characteristics of the input networks.

\begin{assumption} \label{assump:exploration_nodes}
    There exist $d$ edges $e^{\circ}_i$, $1\leq i\leq d$ such that $X = [x_{e_1^o},\cdots,x_{e_d^o}] $ is non-singular, with minimum singular value $\sigma_{\min}^o := \sigma_d(X) > 0$  and minimum eigenvalue $\lambda_{\min}^o := \lambda_{\min} \Big ( \sum_{i=1}^d x_{e_i^o} x_{e_i^o}^\top \Big ) > 0$.
\end{assumption}
A singular value decomposition is a generalization of an eigenvalue decomposition.  Every $d \times d$ matrix $X$ can always be decomposed as $X = U \Sigma V^T$, where both $U, V$ are unitary matrices and $\Sigma$ is a diagonal matrix with $\sigma_i := \Sigma_{ii} \geq 0$ for all $1 \leq i \leq d$.  The minimum singular value being positive, i.e., $\sigma_{\min}^o >0$, is equivalent to $X$ being invertible, or the columns of $X$ being linearly independent.

The above assumption can be easily satisfied by projecting the features onto a lower dimensional space.  Under this assumption, if we continue to explore the set of diverse edges consistently, the confidence region for our parameters will contract in all directions, so that our estimator of $\theta^*$ at time $t$, namely $\hat \theta_t$, will converge, i.e., $
\hat \theta_t \rightarrow \theta^\ast$ as $t \rightarrow \infty$. 

For any matrix $A \in \RR^{d\times d}$, we define the \emph{spectral norm} induced by Euclidean norm $\| \cdot\|_2$ as 
$$
\| A\|_2 = \sup \{ \|A x\|_2: x\in \RR^d, \| x\|_2 \leq 1\}.
$$
Indeed, for any matrix $A \in \RR^{d \times d}$, the spectral norm $\|A \|_2$ has the same value as its largest singular value $\sigma_{\max}(A)$.
Next, we make the following boundedness assumption on edge features.
\begin{assumption}\label{assump:upper_bound_sum_features}
For all $v \in \cV$ and all $\cB\subseteq \iN(v)$,
there exists a constant $D > 0$ such that $\Big \| \big ( (\sum_{u\in \cB} x_{u,v})(\sum_{u\in \cB} x_{u,v})^\top \big ) \big ( (\sum_{u\in \cB} x_{u,v})(\sum_{u\in \cB} x_{u,v})^\top \big ) \Big \|_2 \leq D$.
\end{assumption}
 As the features and parameters we consider are finite, this assumption is without loss of generality and can be satisfied with the feature space's appropriate scaling.

Furthermore, 
a classical IM oracle takes the influence probability on every single edge as input and greedily adds the node that maximizes marginal expected reward into the seed set $S$. Fortunately, the objective function $f(S)$ is both monotone and submodular with respect to $S$, i.e.,
 $f(S) \leq f(T)$ and $f(T \cup \{v\}) - f(T) \leq f(S \cup \{v\}) - f(S)$ for all $S \subset T$ and $v \notin T$, so that the greedy algorithm returns an  $(1-1/e)$-approximation to the optimum \cite{kempe2003maximizing}.  

As discussed in Section \ref{sec:challenges}, UCB-type algorithms construct an optimistic estimator for every single edge, which is not guaranteed to improve in the node-level setting. Consequently, the reward will not necessarily improve over the learning process. In contrast, if the estimators derived by a specific parameter $\tilde \theta \in \cC_t$, in particular, $p_e(\tilde \theta)$ for all $e \in \cE$, are adopted as the input, we might be able to improve our decisions by collecting more observations. This motivates us to use a more sophisticated design of oracle.

In what follows, we introduce a special $(\alpha,\beta)$-Oracle that returns an $(\alpha,\beta)$-approximation solution to the optimum associated with a set of influence probabilities.  To be specific, consider a graph $\cG$, cardinality constraint $K$, and confidence ellipsoid $\cC$, 
we define
\begin{equation*}
    ( S_{\cC}^*, \theta_{\cC}^* ) = \argmax_{ |S| \leq K, \theta \in \cC} f \big (S,p(\theta) \big )
\end{equation*}
as the optimal seed-parameter pair that yields the highest expected reward among all possible parameters $\theta \in \cC$ with cardinality no greater than $K$. 
We assume access to the following offline approximation oracle:
\begin{assumption}\label{assump:oracle}
Let graph $\cG$, seed-size $K$, and confidence region $\cC$ be the inputs, 
there exists an $(\alpha,\beta)$-Oracle 
which returns a solution $\tilde S$ with cardinality $|\tilde S| \leq K$ such that 
\begin{equation*}
    \PP \Big (f \big (\tilde S,p(\tilde \theta) \big ) \geq \alpha \cdot f \big (S_{\cC}^*, p( \theta_{\cC}^* ) \big ) \Big ) \geq \beta,
\end{equation*}
for some $\tilde \theta \in \cC$, $\alpha ,\beta >0$.
\end{assumption}

Unlike the UCB-type approaches discussed in Section \ref{sec:challenges} that take an optimistic estimator for each single edge as input, the Oracle seeks a seed set $\tilde S$ associated with an influence parameter $\tilde \theta$ that makes the \emph{entire} reward function $ f \big (\tilde S,p(\tilde \theta) \big ) $ an \emph{optimistic} and \emph{well-performing} estimator of the scaled optimal reward. 
To be specific, 
for any returned solution $\tilde S$, we can see there exists at least one $\tilde \theta \in \cC$ such that
\begin{equation*}
    f \big (\tilde S,p(\tilde \theta) \big ) \geq \alpha \cdot f \big (S_{\cC}^*, p( \theta_{\cC}^* ) \big ) \geq  \alpha \cdot f \big (S^*, p( \theta^* ) \big ), 
\end{equation*}
so that $\alpha \cdot f \big (S^*, p( \theta^* ) \big ) $ is optimistically represented by  $f \big (\tilde S,p(\tilde \theta) \big )$. Meanwhile, for any set of edges $b$ (hyper-edge $x_b$) being observed, 
the aggregated diffusion probability $p(b,\tilde \theta)$ is always bounded by its optimistic and pessimistic estimators within confidence region $\cC$, i.e.,
\begin{equation*}
   \min_{\theta \in \cC} p(b,\theta) \leq  p(b,\tilde \theta) \leq \max_{\theta \in \cC} p(b,\theta).
\end{equation*}
Utilizing this property, the incurred loss could be further characterized with the help of the confidence region $\cC$, so that $ f \big (\tilde S,p(\tilde \theta) \big ) $ is well-performing. 

Intuitively, to achieve such an oracle, if $\cC$ consists of a finite number of parameter candidate $\theta$'s, we could implement this oracle by simply running a greedy algorithm under each $\theta \in \cC$. If $\cC$ is continuous, we could obtain an approximate solution by discretizing the confidence region $\cC$. Indeed, a similar PairOracle under the Linear Threshold models could be implemented effectively for Bipartite and Directed Acycle Graphs \citep{li2020online}. As this paper focuses on online learning solutions, we omit  detailed discussion of this offline oracle and assume  access to such an oracle throughout this section.


\subsection{Online Learning Algorithm}
This part addresses the challenges in node-level learning and provides an online learning algorithm that learns influence probabilities and high-reward seed sets simultaneously.   
Our algorithm consists of \emph{exploration} and \emph{exploitation} phases. We use the exploration phase to sample edges within $\cD^o$ to increase the feature diversity of success edges, regardless of our current estimate $\hat \theta_t$.  An \emph{exploration super-round} consists of $d$ separate exploration rounds, each round selecting a single node $v_i^o := \text{head}(e_i^o)$ for every $e_i^o \in \cD^o$. 
On the other hand, we use the exploitation phase to select good seed sets based on our current estimate $\hat \theta_t$, and update our estimator $\hat \theta_t$ based on node-level observations simultaneously.

In what follows, we provide detailed solutions to the challenges discussed in Section \ref{sec:challenges}. 

\textbf{Lower Bound on $\rho_k^*$:} Our first task is to investigate the distribution of $\rho_k^*$. 
Recall that $\rho_n^*$ is the optimal solution to Prob.\eqref{prob:sdp_rho}. 
To further explore the distribution of $\rho_n^*$, we provide Lemma \ref{lemma:psd} to show that $\rho_n^*$ is lower bounded by a positive constant with a certain probability.  This result is essential to our confidence ball analysis. We provide the detailed proof in Appendix Section \ref{app:B1}.

 \begin{lemma} \label{lemma:psd}
  Let $x_1,\cdots, x_n\in \R^d$ be $n$ vectors with $d \ll n$.  Let $Y_1,\cdots,Y_n$ be $n$ independent binary random variables such that $P(Y_i  = 1) = p$ and $P(Y_i = 0) = 1-p$. Suppose $\| (x_i x_i^\top ) (x_i x_i^\top ) \|_2 \leq D$ for $i=1,\cdots, n$, 
  and let $\lambda^* := \lambda_{\min}(\sum_{i=1}^n x_i x_i^\top )$. 
For any $ c \in [0,1/p)$, we have 
\begin{equation} \label{eq:psd_vector_sum}
\sum_{i=1}^n Y_i x_i x_i^\top  \succeq cp \sum_{i=1}^n x_i x_i^\top , 
\end{equation}
with probability at least $1 -  d \exp \Big (\frac{-(1-c)^2p^2{\lambda^*}^2/2}{nD + (1-c)p^2 \lambda^*/3} \Big ) $.
 \end{lemma}
Recall that $p_{\min}: = \min_{e \in \cE} p(e)$. By setting $c = \frac{1}{2}$, the above result suggests that $\rho_n^*$ is lower bounded by $p_{\min}/2$ with probability at least $1 - d \exp \Big (\frac{-0.25p_{\min}^2{\lambda^*}^2/2}{nD + 0.5p_{\min}^2 \lambda^*/3} \Big )$.

\textbf{Tradeoff between Exploration and Exploitation:}
Intuitively, to ensure that the above  probability approaches to 1 as $n \to \infty$, it is required that ${\lambda^*}^2$ should grow faster than $n$.
Recall \eqref{def:V} that $\M_n = \sum_{i=1}^n x_ix_i^\top $, we can see $\lambda^* = \lambda_{\min}(\sum_{i=1}^n x_i x_i^\top ) = \lambda_{\min}(\M_n) $.  In the next result, we rigorously show that when $\lambda^* \geq \Omega (\sqrt{n} \log n)$, the condition $\V_n \succeq p_{\min} \M_n/2$ holds with high probability.
\begin{lemma} \label{lemma:mat_concentration}
Suppose Assumption \ref{assump:upper_bound_sum_features} holds and $ \Omega (\sqrt{n} \log n) \leq \lambda_{\min}(\M_n)$, then we have $\V_n \succeq  0.5 p_{\min} \M_n$ with probability at least $1- \cO(1/n^{2})$. 
\end{lemma}
We provide the detailed proof in  Appendix Section \ref{app:B2}. We note that $\lambda^* = \lambda_{\min}(\M_n) $ depends on the number of observed hyper-edges $n$ and their features.
Motivated by this result, to ensure that  $\lambda_{\min}(\M_n) \geq \Omega (\sqrt{n} \log n) $, we could conduct exploration to improve the diversity of the observed features. 
The following lemma characterizes the relationship between $\lambda_{\min}(\M_n)$ and the number of exploration super-rounds. 
\begin{lemma}\label{lemma:eig}
Suppose Assumption \ref{assump:exploration_nodes} holds, and the algorithm has run $\tau$ exploration super-rounds $\tau$.  Then we have 
$$
\lambda_{\min}(\M_\tau) \geq \tau \cdot \lambda_{\min}^o.
$$
\end{lemma}
Due to space limitation, the detailed proof is deferred to Appendix Section~\ref{app:B3}.
A corollary for the number of exploration rounds $\tau$ directly follows Lemma \ref{lemma:mat_concentration} and \ref{lemma:eig}. 
\begin{corollary}\label{corollary:1}
Suppose the algorithm has conducted $ \tau =  \Omega(\sqrt{T} \log T)$ exploration super-rounds and $T- d \tau $ exploitation rounds, then for all $d\tau \leq t \leq T$, we have $\V_t \succeq 0.5 p_{\min} \M_t $ with probability at least $1- \cO(\frac{1}{T^{\log T}})$.
\end{corollary}

\textbf{The Algorithm:} 
In what follows,
we carefully balance the trade-off between exploration and exploitation,  and propose an online learning algorithm. We partition the time horizon into
two phases, exploration and exploitation. In the exploration phase, we conduct $\tau$ exploration super-rounds ($\tau$ single exploration rounds for each $ x_{e_j^o} \in [x_{e_1^o},\cdots,x_{e_d^o}] $ ). Using Corollary \ref{corollary:1}, we can see that the lower bound, $\V_t \succeq 0.5 p_{\min} \M_t $, is achieved  with high probability under the choice $\tau = \sqrt{T}\log T$. We then conduct an exploitation phase in the remaining $T - d\tau$ rounds.  In this phase, we  iteratively invoke the Oracle, and update our estimators $\hat \theta_t$ as well as the confidence region $\cC_t$ by using node-level observations.

The complete Two-Phase Node-level Influence Maximization (TPNodeIM) algorithm is summarized in Algorithm \ref{alg:IM01}. Note that by contrast with the LT-LinUCB algorithm of \cite{li2020online} for a  node-level Linear Threshold model that performs exploitation alone, our algorithm requires unique treatment to ensure that the value $\rho_t^*$ returned by the SDP \eqref{prob:sdp_rho} is lower bounded by a constant.

\begin{algorithm}[t]
 \caption{Two-Phase Node-level Influence Maximization (TPNodeIM)} \label{alg:IM01}
\begin{algorithmic}
\STATE{\bfseries Input:}{ graph $\cG$, $(\alpha,\beta)$-Oracle, feature vector $x_e$'s, total rounds $T$, tuning parameter $\tau$.}
\STATE{\bfseries Initialization:}{ round $t \leftarrow 0$, $\V_0 \leftarrow 0_{d\times d}$, $\M_0 \leftarrow 0_{d\times d}$. }
\STATE{\bfseries Exploration Phase:}
\FOR{ $k= 1,2,\cdots, \tau$} 
\STATE{Run exploration super-rounds for each edge $ x_{e_j^o} \in [x_{e_1^o},\cdots,x_{e_d^o}] $.}
\ENDFOR
\STATE{
Observe feedback in past rounds.  Update $
 \V_{d\tau} = \sum_{i=1}^{d\tau}  x_i x_i^\top  Y_i $ and $\M_{d\tau}  = \sum_{i=1}^{d\tau}  x_i x_i^\top $ 
by \eqref{def:V}. 
}
\STATE{ 
Derive MLE $\hat \theta_{d\tau }$ by \eqref{eq:gradient} 
   $  \nabla l(\hat \theta_{d\tau}) = 0$, solve SDP \eqref{prob:sdp_rho} $\rho_{d\tau}^* = \max \{ \rho | \V_{d\tau} - \rho \M_{d\tau} \succeq 0 \} $.
}

\STATE{
Construct confidence ellipsoid \eqref{eq:confidence_ball_1} using $\V_{d\tau}, \M_{d\tau}, \rho^*_{d\tau}$ and $\hat \theta_{d\tau}$. 
}
\STATE{\bfseries Exploitation Phase:}
\FOR{$t=d\tau + 1,\cdots, T$}
\STATE{Set $S_t \in Oracle(\cG,K,\cC_{t-1})$ as the seed set. }
\STATE{Observe node-level feedback, update $\V_t$, $\M_t$, solve $ \hat \theta_t, \rho_t^*$.}
\STATE{Construct confidence ellipsoid $\cC_t$ by 
$\V_{t}, \M_t, \rho^*_t$ and $\hat \theta_t$.
}
\ENDFOR
 \end{algorithmic}
\end{algorithm}
\subsection{Terminology}
We introduce some terminology to investigate important network characteristics better. 
In a directed graph $\cG = (\cV, \cE)$, given any seed set $S \subset \cV$, for any node $v \in \cV\setminus S$, we denote by $\cG_{S,v} = (\cV_{S,v}, \cE_{S,v})$ the subgraph consisting of directed paths from any seed node $s\in \cS$ to $v$.
Meanwhile, a set of edges $b \subset \cE_{S,v}$ is said to be \emph{relevant} to $v$ if every edge $e\in b$ shares a same follower node $u \in \cV_{S,v}$, and we denote by $\cB_{S,v}$ the collection of all relevant edge sets. 

Next we consider the topology of graph $\cG$.  For any given edge set $b \in \cB_{S,v}$, we denote by 
 $N_{S,b} :=  \sum_{v \in \cV \setminus S} \1 \{b \text{ is relevant to } v \text{ under } S\}$ the number of nodes that $b$ is relevant to, and denote by $P_{S,b} := \PP(\,\, b \text{ is observed }|\,\, S)$ the probability that the edge set $b$ is observed. Also, we denote by $\cB_{S}: = \cup_{v \in \cV} \cB_{S,v}$ the collection of all relevant edge sets under $S$, denote by $G_{S} :=  \sum_{b \in \cB_{S}} P_{S,b}\cdot N_{S,b}^2$, and denote by $G^* : = \max_{|S| \leq K } G_S$.

\subsection{Regret Analysis} 
We analyze the cumulative regret of Algorithm \ref{alg:IM01}, which is defined to be the cumulative loss in reward compared to the optimal policy. The regret is incurred by  both inaccurate estimation of diffusion probability and the randomness involved in the $(\alpha,\beta)$-oracle. To be specific, the oracle only guarantees an $\alpha$-approximation with probability $\beta$, which results in an inevitable loss if the optimal solution is set as the benchmark. Thus, we adopt the $(\alpha,\beta)$-scaled regret proposed in \cite{wen2017online} as our formal metric. Let $f^* = f \big (S^*, p( \theta^* )$ be the optimal expected influence nodes, for any seed set $S$, following this benchmark, we define $R^{\alpha \beta} (S) := f^* - \frac{1}{\alpha \beta} \EE[ f \big (S, p(\theta^*) \big )] $ as the expected \emph{$(\alpha,\beta)$-scaled regret} with seed set $S$ and  $\theta^*$, and denote by $R_t^{\alpha \beta} := R^{\alpha \beta}(S_t)$ the $(\alpha,\beta)$-scaled regret incurred in round $t$ by running Algorithm \ref{alg:IM01}. 
Note that this $(\alpha,\beta)$-scaled regret reduces to the standard regret with the existence of a more potent offline oracle for which $\alpha = \beta = 1$. We have the following theorem providing a guarantee on the performance of Algorithm \ref{alg:IM01}.
\begin{theorem}
 Suppose Assumptions \ref{assumption:feature_norm}, \ref{assump:exploration_nodes}, \ref{assump:upper_bound_sum_features},
 and \ref{assump:oracle} hold. Let $\kappa >0$ be a constant satisfying \eqref{def:kappa}
, let $R := \max_{e \in \cE} 1/p(e)$, and let $\rho = p_{\min}/2 = \min_{e \in \cE} p(e)/2 >0$.
  Let  $\tau = \max \Big ( \sqrt{T} \log T, 16R^2(d +2 \log(T) )/(\kappa^2 \rho^2 \lambda_{\min}^o)  \Big ) $, let $c =\sqrt{ \frac{4R^2}{\kappa^2 \rho } \Big ( \frac{d}{2} \log(1 + 2T |\cE|/d) +2 \log (T)\Big )}$, and let Algorithm \ref{alg:IM01} run $T$ rounds.
 Then we have
 \begin{equation*}
     \sum_{t=1}^T R_t^{\alpha \beta}\leq d\tau f^*+ \frac{2c}{\sqrt{\rho} \alpha \beta} \cdot  \sqrt{(T - d \tau)  G^*} \sqrt{  2d |\cE|  \log(1+\frac{(T-d\tau) |\cE|}{d})} +  \cO \big ((f^*-K)\pi^2/3 \big ). 
 \end{equation*}
 \end{theorem}
\textit{Proof.}
Recall that our algorithm first conducts $\tau =\max \Big ( \sqrt{T} \log T, 16R^2(d +2\log(T) )/(\kappa^2 \rho^2 \lambda_{\min}^o)  \Big ) $ exploration super-rounds ($d\tau$ exploration rounds) with $\rho = 0.5 p_{\min}$.  For any exploitation round $t \geq d\tau +1$, we define the favorable event $\zeta_t$ as $$\zeta_t := \ind \{ \V_t \succeq 0.5p_{\min} \cdot \M_t\}.$$
By Corollary \ref{corollary:1}, we have that  $\PP(\bar \zeta_t)  \leq \cO(1/t^2)$ for every exploitation round $t=  d \tau+1,  d\tau+ 2,\cdots, T$. 

Meanwhile, we denote by $n_t^{\text{ob}}$ the total number of observed hyper-edges up to round $t$. Under event $\zeta_t$, 
by setting $\delta_1 = \delta_2 = 1/t^2$, we have $\rho^2\kappa^2 \lambda_{\min}(\M_t) \geq \rho^2\kappa^2 \tau \lambda_{\min}^o \geq   16R^2\Big ( d + 2\log(t) \Big ) $, so that the condition in Theorem \ref{thm:confidence_ball} is satisfied. By Theorem \ref{thm:confidence_ball}, we obtain that $\theta^* $ falls in the confidence ball
$$
  \cC_t : = \left \{ \theta: (\theta - \theta_t)^\top  \V_t (\theta - \theta_t) \leq \frac{4R^2}{\kappa^2 \rho} \Big ( \frac{d}{2} \log(1 + 2n_t^{\text{ob}}/d) + 2\log (t)\Big )   \right \} 
$$
with probability at least $1-2/t^2$. Next, for any exploitation round $t = d\tau +1,\cdots,  T$, we define a more restrict favorable event $\xi_{k}$ as
\begin{equation*}
    \xi_{t}:= \ind \left\{ \theta^* \in \cC_t, \V_t \succeq 0.5 p_{\min} \cdot \M_t \right\} = \ind \left\{ \theta^* \in \cC_t \right \} \cap \zeta_t , \label{good_event}
\end{equation*}
and define $\bar \xi_{t}$ as its complement. Then we have 
\begin{equation}\label{eq:xi}
\PP(\bar \xi_{t}) =  \PP(\theta^* \notin \cC_t | \zeta_t) \PP(\zeta_t) + \PP(\bar \zeta_t) \leq \PP(\theta^* \notin \cC_t | \zeta_t) + \cO(1/t^2) \leq 2/t^2 + \cO(1/t^2) = \cO(1/t^2). 
\end{equation}
In an exploitation round $t$, the regret $R_t^{\alpha \beta}$ incurred in this round is upper bounded by $\EE[  f(S^{opt},\bar p) - \frac{1}{\alpha \beta} (S_t, p( \theta^* )]$ if  $\xi_{t-1}$ holds. Otherwise, since a set of size $K$ is chosen as the seed, the regret $R_t^{\alpha \beta}$ is upper bounded by $(f^*-K)$. In summary, for any exploitation round $t = d\tau +1,\cdots, T$, we have 
\begin{equation}\label{eq:regret_k}
\EE[R^{\alpha \beta}(S_t)]   \leq 
\PP(\xi_{t-1})  \EE \Big [  f\big (S^*, p(\theta^*) \big )  - \frac{1}{\alpha \beta} f\big (S_t, p(\theta^*) \big )  \Big | \xi_{t-1} \Big ] + \PP(\bar \xi_{t-1}) [f^*-K]. 
\end{equation}
Suppose $\xi_{t-1}$ holds, recall that $S_t$ is the solution returned by the $(\alpha,\beta)$-Oracle, then there exists at least one $\tilde \theta_t \in \cC_t$ such that with probability at least $\beta$,
$$
f\big (S_t, p(\tilde \theta_t )\big ) \geq \alpha \cdot \max_{|S|\leq K, \theta \in \cC_t}  f \big (S, p(\theta)\big ) \geq  \alpha \cdot f\big (S^*, p(\theta^*) \big ),
$$
where the last inequality holds by the fact that $\theta^* \in \cC_t$ under $\xi_{t-1}$. This further implies 
$$
f(S^*, p( \theta^*)) \leq \frac{1}{\alpha}f \big (S_t, p( \tilde \theta_t ) \big ) \leq \frac{1}{\alpha \beta} \EE \Big [ f \big (S_t, p( \tilde \theta_t ) \big ) \Big | \xi_{t-1}\Big ].
$$
Plugging the above inequality into  \eqref{eq:regret_k}, we have 
\begin{equation}\label{eq:reg_decompose}
\EE[R_t^{\alpha \beta}]   \leq 
\frac{\PP(\xi_{t-1})}{\alpha \beta} \cdot  \EE \Big [  f \big (S_t, p( \tilde \theta_t) \big ) - f\big (S_t,p( \theta^*) \big ) \Big | \xi_{t-1} \Big ] + \PP(\bar \xi_{t-1}) [f^*-K].
\end{equation}
Furthermore, we denote by $O_t(b)$ the event that a set of edges $b \in \cB_{S_t}$ is observed in round $t$. Then, we can see that every such edge set $b$ corresponds to a hyper-edge with feature $x_b = \sum_{e \in b} x_e$ and activation probability $p(b,\theta^*) = 1-\exp(-x_b^\top \theta^*)$. 
Since we only have access to node-level feedback observations, given a seed set $S_t$, we characterize the differences between the expected number of nodes influenced under $\tilde \theta_t$ with that under the true diffusion parameter $\theta^*$ as follows:
\begin{lemma}\label{lemma:node_level_diff}
For any $t$, any filtration $\cF_{t-1}$, and seed set $S_t$ such that $\xi_{t-1}$ holds, we have 
\begin{equation*}
    \Big | f \big (S_t, p( \tilde \theta_t) \big ) - f\big (S_t,p( \theta^*) \big ) \Big | \leq  \sum_{v\in\cV}  \sum_{b \in \cB_{S_t, v} } \EE \left [\ind \{O_t(b)\} \cdot |p (b, \tilde \theta_t) - p(b,\theta^*)| \Big| \cF_{t-1}, S_t \right ],
\end{equation*}
where $\cB_{S_t,v}$ is the set of relevance subgraphs $\cG_{S_t,v}$. 
\end{lemma}
The details of this proof are provided in Appendix Section \ref{app:B4}.
Meanwhile, using the fact that $\tilde \theta_t, \theta^* \in \cC_{t-1}$ under $\xi_{t-1}$, for any observed hyper-edge $x_b$, we have 
\begin{equation*}
|p(x_b,\tilde \theta_t) - p(x_b, \theta^*)|  =|\exp(x_b^\top \tilde \theta_t) - \exp(x_b^\top  \theta^*)| \leq |x_b^\top \tilde \theta_t - x_b^\top  \theta^*|  \leq 2 c \sqrt{x_b^\top  \V_{t-1}^{-1} x_b}  \leq \frac{2c}{\sqrt{\rho}} \sqrt{x_b^\top  \M_{t-1}^{-1} x_b}, 
\end{equation*}
where the last inequality holds by the fact that $\V_t \succeq \rho \cdot \M_t $ under $\xi_{t-1}$. Combining the above inequality with Lemma \ref{lemma:node_level_diff}, we conclude that 
\begin{equation*}
    \begin{split}
      &  \EE \Big [  f\big (S_t,p( \tilde \theta_t) \big ) - f \big (S_t,p(\theta^*) \big ) \Big | \xi_{t-1} \Big ] \leq 
        \EE \Big [ \sum_{v\in \cV\setminus S_t} 
        \sum_{b\in \cB_{S_t,v}} \ind \{O_t(b)\} \cdot |p (b, \tilde \theta_t) - p(b,\theta^*)|  \Big | \xi_{t-1} \Big ]   \\
    &   \quad  \leq  \frac{2c}{\sqrt{\rho} }\EE \Big [
       \sum_{v \in \cV \setminus S_t} \sum_{b\in \cB_{S_t}} \ind \{O_t(b)\} \cdot  
         \sqrt{x_b^\top  \M_{t-1}^{-1} x_b}
        \Big | \xi_{t-1} \Big ] \\
    & \quad     =  \frac{2c}{\sqrt{\rho} }\EE \Big [
    \sum_{b\in \cB_{S_t}} \ind \{O_t(b)\} N_{S_t,b}\cdot  
         \sqrt{x_b^\top  \M_{t-1}^{-1} x_b}
        \Big | \xi_{t-1} \Big ].
    \end{split}
\end{equation*}
Summing the above inequality over all exploitation rounds $t= d\tau+ 1,\cdots, T$, we have 
\begin{equation*}
    \begin{split}
       & \sum_{t=d\tau + 1}^T \frac{1}{\alpha \beta}\EE \Big [  f \big (S_t, p( \tilde \theta_t) \big ) - f\big (S_t,p( \theta^*) \big ) \Big | \xi_{t-1} \Big ] \\
 & \quad     \leq  \sum_{t=d\tau + 1}^T \frac{2c}{\sqrt{\rho} \alpha \beta}\EE \Big [
        \sum_{b\in \cB_{S_t}} \ind \{O_t(b)\} \cdot   N_{S_t,b}
         \sqrt{x_b^\top  \M_{t-1}^{-1} x_b}
        \Big  | \xi_{t-1}\Big ] 
         \\
     & \quad  \leq  \frac{2c}{\sqrt{\rho}  \alpha \beta} \cdot \EE \left [ \sqrt{  \sum_{t=d\tau +1}^T \sum_{b\in \cB_{S_t}} \ind \{O_t(b)\}  N_{S_t,b}^2 }\cdot
       \sqrt{ \sum_{t=d\tau+1}^T \sum_{b\in \cB_{S_t}} \ind \{O_t(b)\}  x_b^\top  \M_{t-1}^{-1} x_b
        }
        \right ] \\
       & \quad  \leq   \frac{2c}{\sqrt{\rho}  \alpha \beta} \cdot   \sqrt{  \sum_{t=d \tau+1}^T \EE [ \sum_{b\in \cB_{S_t}}  \ind \{O_t(b)\}  N_{S_t,b}^2 ] }\cdot
       \sqrt{  2d |\cE|  \log \big (1+\frac{(T - d \tau )|\cE|}{d}\big )} \\
     & \quad   \leq \frac{2c}{\sqrt{\rho}  \alpha \beta} \cdot  \sqrt{(T - d \tau)  G^*}  \sqrt{  2d |\cE|  \log \big (1+\frac{(T - d \tau )|\cE|}{d} \big )},
    \end{split}
\end{equation*}
where the second last inequality applies standard techniques in bandits literatures (see Lemma 2 in \cite{wen2017online}) and the fact that at most $|\cE|$ hyper-edges could be observed in every learning round. 
Summing \eqref{eq:reg_decompose} over $t=d\tau + 1,\cdots, T$ and 
applying the above inequality, we conclude 
\begin{equation*}
\begin{split}
\sum_{t=d\tau+1}^T \EE[R_t^{\alpha \beta}]  & \leq 
\sum_{t=d\tau + 1} ^T \frac{\PP(\xi_{t-1})}{\alpha \beta} \cdot  \EE\Big [  f \big (S_t, p( \tilde \theta_t) \big ) - f\big (S_t,p( \theta^*) \big ) \Big | \xi_{t-1} \Big ]  + \sum_{t=d\tau + 1} ^T \PP(\bar \xi_{t-1}) [f^*-K]  \\
& \leq  \frac{2c}{\sqrt{\rho}  \alpha \beta} \cdot  \sqrt{(T - d \tau)  G^*} \cdot   \sqrt{  2d |\cE|  \log \big (1+\frac{(T - d\tau) |\cE|}{d} \big )} + \sum_{t=d\tau + 1} ^T \PP(\bar \xi_{t-1})(f^*-K) 
\\
& \leq   \frac{2c}{\sqrt{\rho} \alpha \beta} \cdot  \sqrt{(T - d \tau)  G^* } \cdot   \sqrt{  2d |\cE|  \log \big (1+\frac{(T - d\tau) |\cE|}{d} \big )} + \sum_{t=d\tau + 1} ^T  2(f^*-K)/(t-1)^2 \\
& \leq   \frac{2c}{\sqrt{\rho} \alpha \beta} \cdot  \sqrt{(T - d \tau)  G^*} \cdot   \sqrt{  2d |\cE|  \log \big (1+\frac{(T - d\tau) |\cE|}{d}\big )} +  \cO \big ((f^*-K)\pi^2/3 \big ),
\end{split}
\end{equation*}
where the second inequality uses  \eqref{eq:xi} that $\PP(\bar \xi_{t})\leq \cO(1/t^2)$ and the last inequality holds by the fact that $\sum_{t=1}^\infty 1/t^2 = \pi^2/6$. 

Finally, since the regret incurred in exploration rounds $t=1,\cdots, d\tau$ is at most $f^*$, we obtain
\begin{equation*}
    \begin{split}
       & \sum_{t=1}^T R_t^{\alpha \beta} =  \sum_{t=1}^{d \tau} f^* + \sum_{t=d\tau+1}^T \EE[R_t^{\alpha \beta} ]  \\
     & \quad     \leq  d\tau \cdot f^* + \frac{2c}{\sqrt{\rho}  \alpha \beta} \cdot  \sqrt{(T - d \tau)  G^*} \cdot  \sqrt{  2d |\cE|  \log \big (1+\frac{(T - d\tau) |\cE|}{d} \big )} +  \cO\big ((f^*-K)\pi^2/3 \big ),
    \end{split}
\end{equation*}
which completes the proof. 
\QED

The above theorem implies that our algorithm achieves $\tilde \cO( \sqrt{T})$ cumulative regret for node-level IC models, which matches the regret achieved under edge-level models despite having less information available. To the best of our knowledge, this is the first problem-independent bound for node-level IC models.
Meanwhile, we note that a comparable $\tilde \cO(\sqrt{T})$ cumulative regret for node-level LT models is achieved by LT-LinUCB of \cite{li2020online}. Compared with LT-LinUCB, due to the unique inherent likelihood function for node-level IC models, our TPNodeIM algorithm conducts extra $\Theta( \sqrt{T} \log(T))$ exploration rounds to increase the diversity of observed features and further ensure $\V_t \succeq 0.5 p_{\min} \M_t$ throughout the exploitation phase. Furthermore, we point out that both exploration and exploitation phases of TPNodeIM induce regrets of $\tilde \cO(\sqrt{T})$, which implies that the 
exploration phase does not adversely affect the total regret along the time horizon $T$.

\section{Conclusion}\label{sec:conclusion}
In this paper, we study the online influence maximization under node-level feedback Independent Cascade models, where only the status of nodes instead of edges are observed. This model finds widespread applications in investigating complex real-world networks but lacks analysis due to a few challenges in constructing confidence ellipsoid and designing online algorithms. We build up the confidence ellipsoid and develop a two-phase learning algorithm TPNodeIM that effectively learns the influence probabilities and achieves $\tilde \cO(\sqrt{T})$ regret.

\bibliographystyle{informs2014} 
\bibliography{source.bib}

\begin{thebibliography}{39}
\providecommand{\natexlab}[1]{#1}
\providecommand{\url}[1]{\texttt{#1}}
\providecommand{\urlprefix}{URL }

\bibitem[{Abbasi-Yadkori et~al.(2011)Abbasi-Yadkori, P{\'a}l, \protect\BIBand{}
  Szepesv{\'a}ri}]{abbasi2011improved}
Abbasi-Yadkori Y, P{\'a}l D, Szepesv{\'a}ri C (2011) Improved algorithms for
  linear stochastic bandits. \emph{Advances in Neural Information Processing
  Systems}, 2312--2320.

\bibitem[{Agrawal et~al.(2017)Agrawal, Avadhanula, Goyal, \protect\BIBand{}
  Zeevi}]{agrawal2017thompson}
Agrawal S, Avadhanula V, Goyal V, Zeevi A (2017) Thompson sampling for the
  mnl-bandit. \emph{Conference on Learning Theory}, 76--78 (PMLR).

\bibitem[{Agrawal et~al.(2019)Agrawal, Avadhanula, Goyal, \protect\BIBand{}
  Zeevi}]{agrawal2019mnl}
Agrawal S, Avadhanula V, Goyal V, Zeevi A (2019) Mnl-bandit: A dynamic learning
  approach to assortment selection. \emph{Operations Research}
  67(5):1453--1485.

\bibitem[{Audibert et~al.(2014)Audibert, Bubeck, \protect\BIBand{}
  Lugosi}]{audibert2014regret}
Audibert JY, Bubeck S, Lugosi G (2014) Regret in online combinatorial
  optimization. \emph{Mathematics of Operations Research} 39(1):31--45.

\bibitem[{Bharathi et~al.(2007)Bharathi, Kempe, \protect\BIBand{}
  Salek}]{Bharathi2007competitive}
Bharathi S, Kempe D, Salek M (2007) Competitive influence maximization in
  social networks. \emph{Proceedings of the 3rd International Conference on
  Internet and Network Economics}, 306–311, WINE'07 (Berlin, Heidelberg:
  Springer-Verlag), ISBN 3540771042.

\bibitem[{Chen et~al.(2020)Chen, Li, \protect\BIBand{} Yang}]{chen2020revenue}
Chen N, Li A, Yang S (2020) Revenue maximization and learning in products
  ranking. \emph{arXiv preprint arXiv:2012.03800} .

\bibitem[{Chen et~al.(2016{\natexlab{a}})Chen, Lin, Tan, Zhao,
  \protect\BIBand{} Zhou}]{Chen2016RobustIM}
Chen W, Lin T, Tan Z, Zhao M, Zhou X (2016{\natexlab{a}}) Robust influence
  maximization. \emph{Proceedings of the 22nd ACM SIGKDD International
  Conference on Knowledge Discovery and Data Mining}, 795–804, KDD '16 (New
  York, NY, USA: Association for Computing Machinery), ISBN 9781450342322.

\bibitem[{Chen et~al.(2013)Chen, Wang, \protect\BIBand{}
  Yuan}]{chen2013combinatorial}
Chen W, Wang Y, Yuan Y (2013) Combinatorial multi-armed bandit: General
  framework and applications. \emph{International Conference on Machine
  Learning}, 151--159 (PMLR).

\bibitem[{Chen et~al.(2016{\natexlab{b}})Chen, Wang, Yuan, \protect\BIBand{}
  Wang}]{chen2016combinatorial}
Chen W, Wang Y, Yuan Y, Wang Q (2016{\natexlab{b}}) Combinatorial multi-armed
  bandit and its extension to probabilistically triggered arms. \emph{The
  Journal of Machine Learning Research} 17(1):1746--1778.

\bibitem[{Cheung et~al.(2019)Cheung, Tan, \protect\BIBand{}
  Zhong}]{cheung2019thompson}
Cheung WC, Tan V, Zhong Z (2019) A thompson sampling algorithm for cascading
  bandits. \emph{The 22nd International Conference on Artificial Intelligence
  and Statistics}, 438--447.

\bibitem[{Combes et~al.(2015)Combes, Talebi, Proutiere, \protect\BIBand{}
  Lelarge}]{combes2015combinatorial}
Combes R, Talebi MS, Proutiere A, Lelarge M (2015) Combinatorial bandits
  revisited. \emph{Proceedings of the 28th International Conference on Neural
  Information Processing Systems - Volume 2}, 2116–2124, NIPS'15 (Cambridge,
  MA, USA: MIT Press).

\bibitem[{Fisher(1997)}]{Fisher001OA}
Fisher R (1997) On an absolute criterion for fitting frequency curves.

\bibitem[{Gai et~al.(2012)Gai, Krishnamachari, \protect\BIBand{}
  Jain}]{gai2012combinatorial}
Gai Y, Krishnamachari B, Jain R (2012) Combinatorial network optimization with
  unknown variables: Multi-armed bandits with linear rewards and individual
  observations. \emph{IEEE/ACM Transactions on Networking} 20(5):1466--1478.

\bibitem[{Goyal et~al.(2010)Goyal, Bonchi, \protect\BIBand{}
  Lakshmanan}]{goyal2010learning}
Goyal A, Bonchi F, Lakshmanan LV (2010) Learning influence probabilities in
  social networks. \emph{Proceedings of the third ACM international conference
  on Web search and data mining}, 241--250 (ACM).

\bibitem[{He \protect\BIBand{} Kempe(2016)}]{He2016RobustIM}
He X, Kempe D (2016) Robust influence maximization. \emph{Proceedings of the
  22nd ACM SIGKDD International Conference on Knowledge Discovery and Data
  Mining} .

\bibitem[{He \protect\BIBand{} Kempe(2018)}]{He2018RobustIM}
He X, Kempe D (2018) Stability and robustness in influence maximization 12(6),
  ISSN 1556-4681.

\bibitem[{Katariya et~al.(2016)Katariya, Kveton, Szepesvari, \protect\BIBand{}
  Wen}]{katariya2016dcm}
Katariya S, Kveton B, Szepesvari C, Wen Z (2016) Dcm bandits: Learning to rank
  with multiple clicks. \emph{International Conference on Machine Learning},
  1215--1224.

\bibitem[{Kempe et~al.(2003)Kempe, Kleinberg, \protect\BIBand{}
  Tardos}]{kempe2003maximizing}
Kempe D, Kleinberg J, Tardos {\'E} (2003) Maximizing the spread of influence
  through a social network. \emph{Proceedings of the ninth ACM SIGKDD
  international conference on Knowledge discovery and data mining}, 137--146
  (ACM).

\bibitem[{Kveton et~al.(2015{\natexlab{a}})Kveton, Szepesvari, Wen,
  \protect\BIBand{} Ashkan}]{kveton2015cascading}
Kveton B, Szepesvari C, Wen Z, Ashkan A (2015{\natexlab{a}}) Cascading bandits:
  Learning to rank in the cascade model. \emph{International Conference on
  Machine Learning}, 767--776.

\bibitem[{Kveton et~al.(2015{\natexlab{b}})Kveton, Wen, Ashkan,
  \protect\BIBand{} Szepesvari}]{kveton2015combinatorial}
Kveton B, Wen Z, Ashkan A, Szepesvari C (2015{\natexlab{b}}) Combinatorial
  cascading bandits. \emph{Advances in Neural Information Processing Systems},
  1450--1458.

\bibitem[{Kveton et~al.(2015{\natexlab{c}})Kveton, Wen, Ashkan,
  \protect\BIBand{} Szepesvari}]{kveton2015tight}
Kveton B, Wen Z, Ashkan A, Szepesvari C (2015{\natexlab{c}}) Tight regret
  bounds for stochastic combinatorial semi-bandits. \emph{Artificial
  Intelligence and Statistics}, 535--543 (PMLR).

\bibitem[{Lei et~al.(2015)Lei, Maniu, Mo, Cheng, \protect\BIBand{}
  Senellart}]{OIM}
Lei S, Maniu S, Mo L, Cheng R, Senellart P (2015) Online influence
  maximization. \emph{Proceedings of the 21th ACM SIGKDD International
  Conference on Knowledge Discovery and Data Mining}, 645--654, KDD '15 (New
  York, NY, USA: ACM), ISBN 978-1-4503-3664-2.

\bibitem[{Li et~al.(2017)Li, Lu, \protect\BIBand{} Zhou}]{li2017provably}
Li L, Lu Y, Zhou D (2017) Provably optimal algorithms for generalized linear
  contextual bandits. \emph{Proceedings of the 34th International Conference on
  Machine Learning-Volume 70}, 2071--2080 (JMLR. org).

\bibitem[{Li et~al.(2020)Li, Kong, Tang, Li, \protect\BIBand{}
  Chen}]{li2020online}
Li S, Kong F, Tang K, Li Q, Chen W (2020) Online influence maximization under
  linear threshold model. \emph{Proceedings of the 33rd Conference on Neural
  Information Processing Systems (NeurIPS)} .

\bibitem[{Mannor \protect\BIBand{} Shamir(2011)}]{mannor2011bandits}
Mannor S, Shamir O (2011) From bandits to experts: On the value of
  side-observations. \emph{Advances in Neural Information Processing Systems},
  volume~24.

\bibitem[{McCullagh(2019)}]{mccullagh2019generalized}
McCullagh P (2019) Generalized linear models .

\bibitem[{Nemhauser et~al.(1978)Nemhauser, Wolsey, \protect\BIBand{}
  Fisher}]{Nemhauser1978}
Nemhauser GL, Wolsey LA, Fisher ML (1978) An analysis of approximations for
  maximizing submodular set functions i. \emph{Mathematical Programming}
  14(1):265--294, ISSN 1436-4646.

\bibitem[{Netrapalli \protect\BIBand{} Sanghavi(2012)}]{netrapalli2012learning}
Netrapalli P, Sanghavi S (2012) Learning the graph of epidemic cascades.
  \emph{ACM SIGMETRICS Performance Evaluation Review} 40:211--222.

\bibitem[{Oh \protect\BIBand{} Iyengar(2021)}]{oh2021multinomial}
Oh Mh, Iyengar G (2021) Multinomial logit contextual bandits: Provable
  optimality and practicality. \emph{arXiv preprint arXiv:2103.13929} .

\bibitem[{Saito et~al.(2008)Saito, Nakano, \protect\BIBand{}
  Kimura}]{saito2008prediction}
Saito K, Nakano R, Kimura M (2008) Prediction of information diffusion
  probabilities for independent cascade model. \emph{International Conference
  on Knowledge-Dased and Intelligent Information and Engineering Systems},
  67--75 (Springer).

\bibitem[{Tropp et~al.(2015)}]{tropp2015introduction}
Tropp JA, et~al. (2015) An introduction to matrix concentration inequalities.
  \emph{Foundations and Trends{\textregistered} in Machine Learning}
  8(1-2):1--230.

\bibitem[{Valko(2016)}]{valko2016bandits}
Valko M (2016) \emph{Bandits on graphs and structures}. Ph.D. thesis.

\bibitem[{Vaswani \protect\BIBand{} Duttachoudhury(2013)}]{vaswanilearning}
Vaswani S, Duttachoudhury N (2013) Learning influence diffusion probabilities
  under the linear threshold model .

\bibitem[{Vaswani et~al.(2015)Vaswani, Lakshmanan, Schmidt
  et~al.}]{vaswani2015influence}
Vaswani S, Lakshmanan L, Schmidt M, et~al. (2015) Influence maximization with
  bandits. \emph{arXiv preprint arXiv:1503.00024} .

\bibitem[{Wang \protect\BIBand{} Chen(2017)}]{Wang2017ImprovingRB}
Wang Q, Chen W (2017) Improving regret bounds for combinatorial semi-bandits
  with probabilistically triggered arms and its applications. \emph{NIPS}.

\bibitem[{Wen et~al.(2017)Wen, Kveton, Valko, \protect\BIBand{}
  Vaswani}]{wen2017online}
Wen Z, Kveton B, Valko M, Vaswani S (2017) Online influence maximization under
  independent cascade model with semi-bandit feedback. \emph{Advances in Neural
  Information Processing Systems}, 3022--3032.

\bibitem[{Wilkinson(1965)}]{wilkinson1965algebraic}
Wilkinson JH (1965) \emph{The algebraic eigenvalue problem}, volume 662 (Oxford
  Clarendon).

\bibitem[{Yang et~al.(2019)Yang, Wang, \protect\BIBand{}
  Truong}]{yang2019online}
Yang S, Wang S, Truong VA (2019) Online learning and optimization under a new
  linear-threshold model with negative influence. \emph{arXiv preprint
  arXiv:1911.03276} .

\bibitem[{Zong et~al.(2016)Zong, Ni, Sung, Ke, Wen, \protect\BIBand{}
  Kveton}]{zong2016cascading}
Zong S, Ni H, Sung K, Ke NR, Wen Z, Kveton B (2016) Cascading bandits for
  large-scale recommendation problems. \emph{Proceedings of the Thirty-Second
  Conference on Uncertainty in Artificial Intelligence}, 835–844, UAI'16
  (Arlington, Virginia, USA: AUAI Press), ISBN 9780996643115.

\end{thebibliography}

\begin{appendices}

\newpage
\section*{Appendix}
\section{Proofs of Technical Lemmas} \label{app:B}

\subsection{Proof of Lemma \ref{lemma:psd}}\label{app:B1}
 \proof{Proof.}
Denote by $\M_n = \sum_{i=1}^n x_i x_i^\top$. The desired inequality can be equivalently expressed as
\begin{align}\label{eq:sdp_yp}
\sum_{i=1}^n (Y_i -p)x_i x_i^\top \succeq & (c-1) p \M_n.
\end{align}
Let $\lambda^* := \lambda_{\min}(\M_n)$. A sufficient condition for the above inequality to hold is 
$$
\lambda_{\min} \Big (\sum_{i=1}^n (Y_i -p)x_i x_i^\top \Big ) \geq \lambda_{\max} ((c-1)p \M_n) = (c-1)p \cdot\lambda_{\min}(\M_n) = (c-1)p \lambda^*. 
$$
Let $X_i := (Y_i - p)x_ix_i^\top $ for $i=1,\cdots, n$. Since $\EE[Y_i]= p$, it can be seen that $X_i$'s are independent zero-mean symmetric random matrices with $\EE[ X_i] = 0$. Meanwhile, we have $\lambda_{\min}(X_i) = \lambda_{\min}\Big ((Y_i - p) x_i x_i^\top \Big ) \geq \lambda_{\min}(-p x_i x_i^\top ) = -p \|x_i \| \geq -p$, where the last inequality holds by $\|x_i \| \leq 1 $ in Assumption \ref{assumption:feature_norm}. Let $Z := \sum_{i=1}^n X_i$ and define $v(Z)$ as the matrix variance statistics of the sum, i.e., 
\begin{equation*}
    v(Z) = \| \EE[ Z Z] \|_2 = \|\sum_{i=1}^n \EE[ X_i X_i ]\|_2.
\end{equation*} 
Using Assumption \ref{assump:upper_bound_sum_features}
that $\| (x_i x_i^\top)(x_i x_i^\top) \|_2 \leq D$, we have 
 \begin{equation}\label{eq:matrix_ber_01}
   v(Z) \leq \sum_{i=1}^n \| \EE[X_i X_i]\|_2  =  \sum_{i=1}^n \EE[ (Y_i - p)^2] \cdot \| (x_i x_i^\top)(x_i x_i^\top)\|_2 \leq n D. 
 \end{equation}
  By Lemma \ref{lemma:matrix_ber} in Appendix Section \ref{app:B5}, for a finite sequence $\{ X_k\}$ of independent symmetric random matrices of dimension $d\times d$, if $\EE[X_k] = 0$ and $\lambda_{\min}(X_k) \geq -L$ for every $k$, then 
$$
\P\Big (\lambda_{\min} (Z) \leq -t \Big) \leq d \cdot \exp\Big (\frac{-t^2/2}{v(Z) + Lt/3} \Big ),
$$
for any $t \geq 0$. By setting $L = p$, it can be seen that $X_k$'s satisfy all conditions above. Substituting $t = (1-c)p \lambda^*$ and $v(Z) \leq nD$ \eqref{eq:matrix_ber_01} into the preceding equation, we have
\begin{equation*}
    \P (\lambda_{\min}(Z) \leq (c-1) p \lambda^*) \leq d \cdot \exp \Big (\frac{-(1-c)^2p^2{\lambda^*}^2/2}{v(Z) + (1-c)p^2 \lambda^*/3} \Big )  \leq d \cdot \exp \Big (\frac{-(1-c)^2p^2{\lambda^*}^2/2}{nD + (1-c)p^2 \lambda^*/3} \Big )  .
\end{equation*}
Since $\lambda_{\min}(Z) = \lambda_{\min}\Big (\sum_{i=1}^n (Y_i -p)x_i x_i^\top \Big )  \geq (c-1)p\lambda^*$ is a sufficient condition for the desired inequality \eqref{eq:sdp_yp}, we conclude that
\begin{equation*}
    \P \Big (  \sum_{i=1}^n (Y_i - cp)x_i x_i^\top  \succeq  0 \Big  ) \geq 1 - d \cdot \exp \Big (\frac{-(1-c)^2p^2{\lambda^*}^2/2}{nD + (1-c)p^2 \lambda^*/3} \Big ),
\end{equation*}
which completes the proof. 
\QED

\subsection{Proof of Lemma \ref{lemma:mat_concentration}}\label{app:B2}
\proof{Proof.}
Suppose $\lambda_{\min} (\M_n) \geq C \sqrt{n }\log n$ for all $n$, by Lemma \ref{lemma:psd} with the choice of $c = 0.5$, we have 
\begin{align*}
        & \P \Big (  \sum_{i=1}^n (Y_i - 0.5p_{\min})x_i x_i^\top \succeq  0 \Big  ) \\
    &    \quad  \geq  1 - d \exp \left (\frac{-0.25p_{\min}^2{\lambda_{\min}(M_n)}^2/2}{nD + 0.5p_{\min}^2 \lambda_{\min}(M_n)/3} \right ) \\
        & \quad \geq  1- d \cdot  \max  \left \{ \exp \Big (\frac{-0.25p_{\min}^2 \lambda_{\min}^2(M_n) /2}{2nD} \Big ), \exp \Big (\frac{-0.25p_{\min}^2 \lambda_{\min}^2(M_n) /2}{p_{\min}^2 \lambda_{\min}(M_n)/3} \Big )\right \},
\end{align*}
where we use the fact that $\exp(-\frac{c}{a+b}) \leq \max \{ \exp(-\frac{c}{2a}),\exp(-\frac{c}{2b}) \}$ for $a,b,c >0$ in the last inequality. 
First of all, suppose $\exp \Big (\frac{-0.25p_{\min}^2 \lambda_{\min}^2(M_n) /2}{2nD} \Big ) \geq \exp \Big (\frac{-0.25p_{\min}^2 \lambda_{\min}^2(M_n) /2}{p_{\min}^2 \lambda_{\min}(M_n)/3} \Big )$, then we have 
\begin{align*}
\P \Big (  \sum_{i=1}^n (Y_i - 0.5p_{\min})x_i x_i^\top \succeq  0 \Big  ) & \geq  1 - d \exp \left (\frac{-0.25p_{\min}^2C^2 n (\log n )^2/2}{2nD } \right )   \\
& =  1 - d \exp \left (\frac{-0.25p_{\min}^2C^2  (\log n)^2/2}{2D } \right ) \geq 1 - \cO(1/n^{2}),
\end{align*}
where the last inequality holds by the fact that $\exp(-s(\log n )^2) = n^{-s\ln n} \leq \cO(1/n^{2})$ for $s>0$. Secondly, suppose $\exp \Big (\frac{-0.25p_{\min}^2 \lambda_{\min}^2(M_n) /2}{2nD} \Big ) < \exp \Big (\frac{-0.25p_{\min}^2 \lambda_{\min}^2(M_n) /2}{p_{\min}^2 \lambda_{\min}(M_n)/3} \Big )$, then we have 
\begin{align*}
\P \Big (  \sum_{i=1}^n (Y_i - 0.5p_{\min})x_i x_i^\top \succeq  0 \Big  ) & \geq  1 - d \exp \left (\frac{-0.25p_{\min}^2 \lambda_{\min}(M_n)/2}{p_{\min}^2 /3 } \right )  \\
& \geq  1 - d \exp \left (\frac{-0.25p_{\min}^2 C \sqrt{n} \log n /2}{p_{\min}^2 /3 } \right )
\geq 1- \cO(1/n^{2}).
\end{align*}
Combining the above two cases, we have $\V_n \succeq 0.5p_{\min} \M_n$ with probability at least $1- \cO(1/n^{2})$, which completes the proof.
 \QED
 \subsection{Proof of Lemma \ref{lemma:eig}}\label{app:B3}
\proof{Proof.}Consider $\M_{k-1}$, which is the rank-one sum of hyper-edges observed by the end of the $(k-1)$-th exploration super-round.
By definition, we have 
$$
\M_k \succeq \M_{k-1} + \sum_{e_i^o \in \cD} x_{e_i^o} x_{e_i^o}^\top .
$$
That is, we update $\M_{k-1}$ to $\M_k$ by adding together the rank-one sum of hyper-edges from both exploration and exploitation rounds.  By the minimal eigenvalue result in \cite{wilkinson1965algebraic}, for any two positive semi-definite matrices $A$ and $B$, we have 
 $$
 \lambda_{\min}(A) + \lambda_{\min}(B) \leq \lambda_{\min}(A+B). 
 $$
 Thus, we obtain
$$
\lambda_{\min}(\M_k) \geq \lambda_{\min}(\M_{k-1}) + \lambda_{\min}(\sum_{i \in \cD} x_{e_i^o} x_{e_i^o}^\top  ) \geq   \lambda_{\min}(\M_{k-1}) + \lambda_{\min}^o.
$$
Continuing this induction process, since the algorithm has run $k$ exploration super-rounds, we have 
$$
\lambda_{\min}(\M_k) \geq k \cdot \lambda_{\min}^o,
$$
which completes the proof.
\QED

\subsection{Proof of Lemma~\ref{lemma:node_level_diff}}\label{app:B4}
\textit{Proof.}
We denote by 
$f(S, p(\theta),v)$ the probability that node $v$ is activated under seed set $S$ and diffusion probabilities $p(\theta)$. In round $t$, we decompose the difference of expected influenced nodes under $\tilde \theta_t$ and $\theta^*$ in round $t$
as 
\begin{equation*}
f \big (S_t, p(  \tilde \theta_t ) \big ) - f\big (S_t,p( \theta^*) \big ) =  \sum_{v\in\cV\setminus S_t}    \Big ( 
f \big (S_t, p( \tilde \theta_t),v \big ) - f \big (S_t, p( \theta^*),v \big ) \Big )
.
\end{equation*}
It suffices to prove the following result.
\begin{lemma} \label{lemma:node_diff}
Given seed set $S_t$, for any node $v\in \cV \setminus S_t$, we have 
$$| f \big (S_t, p(  \theta),v \big ) - f \big (S_t, p( \theta^*),v \big ) | \leq \sum_{b \in \cB_{S_t, v} } \EE \Big [\ind \{O_t(b)\} \cdot \big |p(b, \theta) - p(b, \theta^*)\big|  \Big |  \cF_{t-1},S_t\Big ].$$
\end{lemma}
We provide the detailed proof in Appendix Section \ref{sec:node_diff}. Then, it follows from Lemma \ref{lemma:node_diff} that
\begin{equation*}
\begin{split}
& | f \big (S_t, p(  \tilde \theta_t ) \big ) - f\big (S_t,p( \theta^*) \big ) | =  \Big  | \sum_{v\in\cV\setminus S_t}    \Big ( 
f \big (S_t, p( \tilde \theta_t),v \big ) - f \big (S_t, p( \theta^*),v \big ) \Big )
\Big |  \\
& \quad \quad \leq  \sum_{v\in\cV\setminus S_t}    \Big | f \big (S_t, p( \tilde \theta_t),v \big ) - f \big (S_t, p( \theta^*),v \big ) \Big  | \\
& \quad  \quad \leq  \sum_{v\in\cV\setminus S_t}  \sum_{b \in \cB_{S_t, v} } \EE \Big [\ind \{O_t(b)\} \cdot \big |p(b, \theta) - p(b, \theta^*)\big|  \Big | \cF_{t-1}, S_t \Big ],
\end{split}
\end{equation*}
which completes the proof. 
\QED
\subsubsection{Proof of Lemma \ref{lemma:node_diff}.}\label{sec:node_diff}
Theorem 3 of \cite{wen2017online} characterized the difference of expected influenced nodes in terms of the estimation error of every observed edge $e\in \cE$. Under node-level feedback, we borrow that idea to bound the difference of expected influenced nodes in terms of estimation error of every observed edge set $b \in \cB_{S_t,v}$. 
We present the details here for completeness. 

\textit{Proof.} Given seed set $S_t$ and node $v \in \cV \setminus S_t$, we consider the diffusion processes on the subgraph $\cG_{S_t,v} = (\cV_{S_t,v}, \cE_{S_t,v})$. For any node $v \in \cV$ and two disjoint sets $\cV_1,\cV_2 \subset \cV_{S_t,v}$,  we denote by $q(\cV_1, \cV_2, p(\theta))$ the probability that $v$ is influenced by $\cV_1$  under diffusion probabilities $p(\theta)$, with $\cV_2$  removed from the subgraph $\cG_{S_t,v}$. 

We denote by $S^s$ the collection of nodes activated at time step $s$, and denote by $\tilde \tau$ the time step that node $v$ is activated or the diffusion terminates. In particular, the diffusion process starts from  seed set $S^0$ at time step 0. Meanwhile, we denote by 
$$
S^{0:k} := \{ u \in S^s, \text{ for } s = 0,1,2,\cdots, k \}
$$
the collection of all activated nodes up to time step $k$. 
Recall that $p(\theta): \cE \rightarrow [0,1] $ are the diffusion probabilities under $\theta$. For time steps $k = 0,1,\cdots, \tilde \tau $, given any $S^k $ and $S^{0:k -1}$, following Lemma 3 of  \cite{wen2017online}, we have
\begin{equation*}
q \big (S^{k}, S^{0:k-1}, p(\theta) \big ) =
 \begin{cases}
 1 & \text{ if }v \in S^k ,  \\
 0 & \text{ if } S^k  = \emptyset,\\
 \EE \big [q \big (S^{k +1}, S^{0:k }, p(\theta) \big )  \big | (S^k , S^{0:k  - 1}) \big ] & \text{ otherwise}.
\end{cases}
\end{equation*}
Clear, when $v\in S^k  $ or $S^k  = \emptyset$, we always have $q(S^k , S^{0:k  -1} , p(\theta)) - q(S^k , S^{0:k  -1} , p(\theta^*)) = 0$. 
Otherwise, given $S^{k }$ and $S^{0:k  - 1}$, we denote by 
$
\nu(S^{k  +1},\theta) \triangleq  \P[S^{k  + 1} | (S^{k }, S^{0,k  - 1}, p(\theta))] 
$
the probability that $S^{k +1}$ is influenced at time step $k  +1$ under $p(\theta)$. Given $S^{0:k }$, consider all possible realizations of $S^{k  +1}$, for all $p(\theta)$, we have 
\begin{align*}
q \big (S^k , S^{0:k  -1}, p(\theta) \big ) = & \sum_{S^{k  +1}} \nu (S^{k  +1 },\theta  ) q \big (S^{k  + 1}, S^{0:k }, p(\theta)  \big ).
\end{align*}
Next, consider $ q \big (S^k , S^{0:k  -1}, p(\theta)  \big ) - q \big (S^k , S^{0:k  -1}, p(\theta^*) \big )  $, by using triangle inequality, we obtain
\begin{equation}\label{eq:1}
\begin{split}
& \Big| q \big (S^k , S^{0:k  -1}, p(\theta)  \big ) - q \big (S^k , S^{0:k  -1}, p(\theta^*) \big )   \Big | \\
& \quad =  \Big |   \sum_{S^{k  +1}}  \Big [\nu  (S^{k +1},\theta ) q \big (S^{k  +1}, S^{0:k }, p(\theta) \big ) - \nu(S^{k  + 1},\theta^*) q \big (S^{k  +1}, S^{0:k }, p(\theta^*) \big ) \Big ]   \Big | \\
& \quad \leq    \Big |   \sum_{S^{k  +1}}  \big [  \nu(S^{k +1},\theta)  - \nu(S^{k  + 1},\theta^*) \big ]\cdot  q \big (S^{k  +1}, S^{0:k }, p(\theta) \big )  \Big |    \\
& \qquad + \sum_{S^{k  + 1}}   \nu (S^{k  + 1},\theta^*) \cdot \Big | q (S^{k  +1}, S^{0:k }, p(\theta)) - q \big (S^{k  +1}, S^{0:k }, p(\theta^*) \big) \Big |  \\
& \quad  =   \Big  |   \sum_{S^{k  +1}}  \big [  \nu (S^{k +1},\theta)  - \nu (S^{k  + 1},\theta^*) \big ]\cdot  q \big (S^{k  +1}, S^{0:k }, p(\theta) \big )  \Big | \\
& \qquad + \EE \Big [ \big | q \big(S^{k  + 1}, S^{0:k  },  p(\theta) \big ) - q \big (S^{k +1}, S^{0:k  },  p(\theta^*) \big )  \big | \Big],
 \end{split}
\end{equation}
where the expectation in the last equality is taken over all possible realizations of $S^{k +1}$. 

Then, we consider $\sum_{S^{k  +1}}  \big [  \nu(S^{k +1},\theta)  - \nu(S^{k  + 1},\theta^*) \big ]\cdot  q \big (S^{k  +1}, S^{0:k }, p(\theta) \big ) $. Let 
\begin{align*}
&    M^+ := \sum_{S^{k  +1}} \big | \nu(S^{k +1},\theta)  - \nu(S^{k  + 1},\theta^*) \big | \cdot \ind  \{ p(S^{k  + 1},\theta ) \geq p(S^{k  + 1},\theta^*)\},  \\
\mbox{ and } &    M^- := \sum_{S^{k  +1}} \big | \nu(S^{k +1},\theta)  - \nu(S^{k  + 1},\theta^*) \big | \cdot  \ind  \{ p(S^{k  + 1},\theta ) < p(S^{k  + 1},\theta^*) \},
\end{align*}
by using the fact that $0 \leq q \big  (S^{k  +1}, S^{0:k }, p(\theta) \big )  \leq 1$, we have 
\begin{equation*}
    \begin{split}
 \Big |   \sum_{S^{k  +1}}  \big [  \nu ( S^{k +1},\theta)  - \nu(S^{k  + 1},\theta^*) \big ]\cdot  q \big (S^{k  +1}, S^{0:k }, p(\theta) \big )  \Big |
 \leq  \max  \{  M^+, M^-\}.
 \end{split}
\end{equation*}
We first note that $\sum_{S^{k  + 1}} \nu (S^{k  + 1},\theta) - \sum_{S^{k  + 1}} \nu (S^{k  + 1},\theta^*) = M^+ - M^- = 0$. Also, we have $M^+ + M^- = \sum_{S^{k  + 1}} |\nu(S^{k  + 1},\theta)- \nu(S^{k  + 1},\theta^*)| $. Clearly, the preceding two equations imply that   $M^+ = M^- = \frac{1}{2} \sum_{S^{k  + 1}} |\nu(S^{k  + 1},\theta)- \nu(S^{k  + 1},\theta^*)|  $. Consequently, the above inequality could be further expressed as 
\begin{equation}\label{eq:node_diff_01}
    \begin{split}
  \Big  |   \sum_{S^{k  +1}}  \big [  \nu (S^{k +1},\theta)  - \nu (S^{k  + 1},\theta^*) \big ]\cdot  q \big (S^{k  +1}, S^{0:k }, p(\theta) \big )  \Big  |
\leq   \frac{1}{2} \sum_{S^{k  + 1}}\big  |\nu (S^{k  + 1},\theta)- \nu (S^{k  + 1},\theta^*) \big |.
 \end{split}
\end{equation}
We denote by  $\cB(S^{k }, S^{0:k  - 1})$ be the collection of edge sets induced by $S^{k }$, with $S^{0:k  - 1}$ being removed from the subgraph $\cG_{S_t,v}$.  By using Lemma \ref{lemma:diff_induction} in Appendix Section \ref{app:diff_induction}, we have 
\begin{equation*} \label{eq:induction_k}
\begin{split}
& \frac{1}{2} \sum_{S^{k  + 1}} |\nu (S^{k  + 1},\theta)- \nu (S^{k  + 1},\theta^*)| \leq   \sum_{b \in  \cB(S^k  , S^{0:k -1})}  | p (b, \theta) - p (b, \theta^*)|   .
\end{split}
\end{equation*}
Substituting the above inequality into \eqref{eq:node_diff_01}, we obtain that 
\begin{equation*} 
 \Big |   \sum_{S^{k  +1}}  \big [  \nu(S^{k +1},\theta)  - \nu (S^{k  + 1},\theta^*) \big ]\cdot  q \big  (S^{k  +1}, S^{0:k }, p(\theta) \big )  \Big |
\leq \sum_{b \in \cB(S^k , S^{0:k -1})} |p(b, \theta) - p(b, \theta^*)|.
\end{equation*}
Combining the above inequality with \eqref{eq:1}, we have
\begin{align*}
 &\Big |q \big (S^k , S^{0:k  -1} , p(\theta) \big ) - q \big (S^k , S^{0:k  -1} , p(\theta^*) \big ) \Big |  \\
& \quad  \leq  \sum_{b \in \cB(S^{k }, S^{0:k  -1})} | p(b,\theta) - p(b,\theta^*)|   + \EE \Big [ \big|q \big (S^{k  + 1}, S^{0:k  } , p(\theta) \big ) - q \big (S^{k +1}, S^{0:k  } , p(\theta^*) \big )\big |    \Big | \cF_{t-1}, S_t \Big ].
\end{align*}
Recall that $\tilde \tau  $ is the stopping time step when the diffusion terminates, by applying the above inequality over time steps $k  = 1,2,\cdots,\tilde \tau$, and following the analysis of Lemma 5 of  \cite{wen2017online}, we obtain 
\begin{equation*}
\begin{split}
&  \big | f \big (S_t, p(\theta), v \big ) - f \big (S_t, p(\theta^*),v \big) \big | =    \big | q \big (S^0, \emptyset, p(\theta) \big ) - q \big (S^0, \emptyset, p(\theta^*)\big )  \big| \\
& \quad \leq  \EE \Big [\sum_{k  = 0}^{\tilde \tau  - 1} \sum_{b \in \cB(S^k , S^{0:k  - 1})}  \big |p(b,\theta) - p(b,\theta^*) \big | \Big ] 
= \EE \Big [ \sum_{b \in \cB_{S_t, v}} \ind \{O_t(b)\}  \cdot  \big |p(b,\theta) - p(b,\theta^*)\big | \Big | \cF_{t-1}, S_t\Big ],
\end{split}
\end{equation*}
where the expectation is taken over the diffusion process under the true probabilities $p(\theta^*)$. This completes the proof. \QED 
\subsubsection{Lemma \ref{lemma:diff_induction} and Its Proof.}\label{app:diff_induction}
\begin{lemma}\label{lemma:diff_induction} 
Given a subgraph $\cG_{S_t,v}$ and time step $k $ of a diffusion process, let $\cB(S^{k }, S^{0:k  - 1})$ be the collection of edge sets induced by $S^{k }$, with $S^{0:k  - 1}$ being removed from the subgraph $\cG_{S_t,v}$, then we have 
\begin{equation*}
\begin{split}
& \frac{1}{2} \sum_{S^{k  + 1}} \big |\nu (S^{k  + 1},\theta)- \nu (S^{k  + 1},\theta^*) \big |  \leq \sum_{b \in  \cB(S^k  , S^{0:k -1})} \big  |p(b, \theta) - p(b, \theta^*) \big |   .
\end{split}
\end{equation*}
\end{lemma}
\textit{Proof.} This result is an extension of Lemma 4 of \cite{wen2017online} to the node-level feedback scenario. We present the detailed proof here for completeness.
Let $\cV(S^k  , S^{0:k -1})$ be the collection of nodes in $\cV \setminus S^{0:k  }$ that can be reached by some nodes in $S^k $ by one directed edge. Then, $\cV(S^k  , S^{0:k -1})$ is the set of candidate nodes that might belong to $S^{k +1}$, and we always have $S^{k  + 1} \subset \cV(S^k  , S^{0:k -1}) $. Let $\tilde v \in \{ 0, 1\}^{\cV(S^k  , S^{0:k -1})}$ be a random realization of the activation status for these candidate nodes, and denote the probability of $\tilde v$ happening under $p(\theta)$ by $\Phi(\tilde v, \theta)$. We can see there exists a one-to-one correspondence between each realization of $\tilde v$ and $S^{k +1}$, which implies that 
\begin{equation}\label{eq:diff_induct_01}
    \frac{1}{2} \sum_{S^{k  + 1}}\big  |\nu (S^{k  + 1},\theta)- \nu (S^{k  + 1},\theta^*) \big | = 
\frac{1}{2} \sum_{ \tilde v \in \{ 0, 1\}^{\cV(S^k  , S^{0:k -1})}} \big |\Phi(\tilde  v, \theta) - \Phi( \tilde v , \theta^*)\big |.
\end{equation}
Meanwhile, 
we can see that the activation status  of each node $v \in \cV(S^k  , S^{0:k -1})$ is uniquely determined by the activation outcome of exactly \emph{one} edge set  $b \in \cB(S^k  , S^{0:k -1})$. Therefore, we can establish a one-to-one correspondence between elements in $\cV(S^k  , S^{0:k -1})$ and $\cB(S^k  , S^{0:k -1})$. 

We index edge sets $b \in \cB(S^{k }, S^{0:k -1})$ as $b_1, ..., b_{|\cB(S^k , S^{0:k -1})|}$.
For any realization $\tilde v$,  we denote by $\tilde v_j$ the realization of the $j$-th node, and denote by $\tilde v_{1:j}$ the realization of the first $j$ nodes. With slight abuse of notations, under influence probability $p(\theta)$, we denote the probability of $\tilde v_j$  happening by $\Phi(\tilde v_{j},\theta)$, and denote the probability of $\tilde v_{1:j}$ happening by $\Phi(\tilde v_{1:j},\theta)$.
Next, we use mathematical induction to prove that 
\begin{equation}\label{eq:diff_induct_02}
\frac{1}{2} \sum_{ \tilde v_{1:j} \in \{ 0, 1\}^{j}} \big |\Phi(\tilde  v_{1:j}, \theta) - \Phi (\tilde v_{1:j},  \theta^*) \big |  \leq  \sum_{i =1}^{j} \big |p(b_i, \theta) - p(b_i, \theta^*) \big |.
\end{equation}
for $j=1,2,\cdots, |\cB(S^k  , S^{0:k -1})|$. 
 
 Consider the case $j =1$, clearly, we have 
\begin{equation*}
\begin{split}
\frac{1}{2} \sum_{ \tilde v_1} \big |\Phi (\tilde v_1,\theta) - \Phi (\tilde v_1,\theta^*) \big | &= \frac{1}{2} \Big \{  \big |p(b_1, \theta) - p(b_1, \theta^*) \big | +  \big |1- p(b_1, \theta) - (1-p(b_1, \theta^*)) \big |\Big \} \\
& = \big |p(b_1, \theta) - p(b_1, \theta^*) \big |.\end{split}
\end{equation*}
To continue the induction process, suppose the hypothesis is true for $j$, in particular, we have
\begin{equation} \label{eq:induction_hyp}
\frac{1}{2} \sum_{ \tilde v_{1:j}  \in \{0,1\}^j} \big |\Phi (\tilde v_{1:j} ,\theta) - \Phi (\tilde v_{1:j} , \theta^*)  \big | \leq \sum_{i =1}^{j} \big |p(b_i, \theta) - p(b_i, \theta^*) \big |.
\end{equation}
We then consider the case $j+1$.  As $\tilde v_{1:j+1}$ is the union of $\tilde v_{1:j}$ and the realization of the $(j+1)$-th node, we have 
\begin{align*}
& \frac{1}{2} \sum_{ \tilde v_{1:j+1}} \big | \Phi (\tilde v_{1:j+1},\theta) - \Phi (\tilde v_{1:j+1},\theta^*) \big | \\
& =  \frac{1}{2} \sum_{\tilde v_{1:j} } \Big \{ \big | \Phi (\tilde v_{1:j},\theta) p(b_{j+1}, \theta) - \Phi (\tilde v_{1:j},\theta^*) p(b_{j+1}, \theta^*) \big |  \\
& \quad + \big | \Phi (\tilde v_{1:j},\theta) \big (1- p(b_{j+1}, \theta) \big ) -  \Phi (\tilde v_{1:j} ,\theta^*) \big (1- p(b_{j+1}, \theta^*) \big) \big | \Big \} \\
& \leq  \frac{1}{2} \sum_{\tilde v_{1:j}}\Big \{ \big | \Phi (\tilde v_{1:j},\theta)  - \Phi (\tilde v_{1:j} ,\theta^*)\big  |  \cdot p(b_{j+1},\theta) +  \Phi (\tilde v_{1:j} ,\theta^*) \cdot \big | p(b_{j+1}, \theta) - p(b_{j+1}, \theta^*) \big |   \\
& \quad + \big | \Phi (\tilde v_{1:j} ,\theta) - \Phi (\tilde v_{1:j},\theta^*) \big | \cdot \big (1- p(b_{j+1}\theta) \big ) + \Phi (\tilde v_{1:j} ,\theta^*) \cdot \big | p(b_{j+1}, \theta)- p(b_{j+1},\theta^*)\big |
\Big \}\\
& =  \frac{1}{2} \sum_{v_{1:j}} \Big \{ \big | \Phi (\tilde v_{1:j} ,\theta)  - \Phi (\tilde v_{1:j} , \theta^*) \big | + 2  \Phi (\tilde v_{1:j} , \theta^*) \cdot \big | p(b_{j+1}, \theta)- p(b_{j+1}, \theta^*)| \Big \}\\
& \leq  \Big ( \sum_{\tilde v_{1:j}} \Phi (\tilde v_{1:j}, \theta^*) \Big ) \cdot \big |p(b_{j+1}, \theta) - p(b_{j+1},\theta^*) \big | + \sum_{i=1}^j \big |p(b_i,\theta) - p(b_i, \theta^*)\big | \\
& =  \sum_{i=1}^{j+1} \big |p(b_i, \theta) - p(b_i, \theta^*) \big |,
\end{align*}
where the second last inequality holds by the induction hypothesis \eqref{eq:induction_hyp}, and the last equation holds by the fact that   $\sum_{\tilde v_j} \Phi  (\tilde v_j,\theta^*) = 1$. This completes the induction procedure. The desired result immediately follows by combining \eqref{eq:diff_induct_01}, \eqref{eq:diff_induct_02}, and the fact that $\sum_{i =1}^{|\cB(S^k  , S^{0:k -1})|} \big |p(b_i, \theta) - p(b_i, \theta^*) \big | =  \sum_{b \in  \cB(S^k  , S^{0:k -1})}  \big |p(b, \theta) - p(b, \theta^*) \big |$. 
\QED

\subsection{Auxiliary Lemmas}\label{app:B5}
Let $\{\cF_t\}_{t=1}^n$ be a sequence of $\sigma$-algebras, and $\{ \epsilon_t\}_{t=1}^n$ be a sequence of mean-zero noises such that $\EE[\epsilon_t | \cF_{t-1} ] = 0$.
Suppose $\epsilon_t$ is conditional $R$-subgaussian such that  
$$
\EE[ e^{\lambda \epsilon_t} ] \leq  e^{\lambda^2 R^2/2}
$$
for all $\lambda$. Let $\M_n = \sum_{t=1}^n x_t x_t^\top $, 
the following two lemmas provide bounds on $\|  \sum_{i=1}^n x_i \epsilon_i\|_{\M_n^{-1}}$. 
\begin{lemma}[Lemma 7, \cite{li2017provably}]\label{lemma:loose_bound}
For any $\delta >0$, with probability at least $1-\delta$, we have 
$$
\|  \sum_{i=1}^n x_i \epsilon_i\|_{\M_n^{-1}} \leq 4R\sqrt{d + \log(1/\delta)}.
$$
\end{lemma}
\begin{lemma}[Lemma 8,  \cite{li2017provably}]\label{lemma:tight_bound}
Suppose there is an integer $m$ such that $\lambda_{\min}(\M_m) \geq 1$, then for any $\delta \in (0,1)$, with probability at least $1-\delta$, for all $t > m$, 
$$
\| \sum_{i=1}^n x_i \epsilon_i\|_{\M_n^{-1}}^2 \leq 4R^2 \Big ( \frac{d}{2} \log(1+ 2n/d) + \log(1/\delta) \Big ).
$$

\end{lemma}

\begin{lemma}[Theorem 6.6.1, \cite{tropp2015introduction}]\label{lemma:matrix_ber} Let $\{X_k\}_{k=1}^n$ be a sequence of random, independent, and symmetric matrices of dimension $d \times d$. Suppose that $\EE[X_k] = 0$ and $\lambda_{\min}(X_k) \geq - L$ for all $k =1,\cdots, n$. Let $Y := \sum_{k=1}^n X_k$, and denote the matrix variance statistics of the sum by $v(Y) $:
$$
v(Y) = \big \| \EE[ Y Y ] \big \|_2 = \big \|\sum_{k=1}^n \EE[X_k X_k] \big \|_2.
$$
Then for all $t \geq 0$, we have $$
\P(\lambda_{\min}(Y) \leq -t) \leq d\cdot \exp \Big( \frac{-t^2/2}{v(Y) + L t/3}  \Big).
$$
\end{lemma}


\end{appendices}

\end{document}